\documentclass[%
unsortedaddress,
preprint,
 amsmath,amssymb,
 aps,
pre
]{revtex4-1}

\usepackage{graphicx}
\usepackage{dcolumn}
\usepackage{bm}

\begin{document}

\title{Shear-induced ordering of nano-pores and instabilities in concentrated surfactant mesh phases}

\author{Pradip K. Bera$^{1}$\footnote[1]{These authors contributed equally to this work.}, Vikram Rathee$^{1}$\footnotemark[1]\footnote[2]{Present address: Okinawa Institute of Science and Technology Graduate University, Okinawa, Japan}, Rema Krishnaswamy$^{1}$\footnote[3]{Present address: School of Arts and Sciences, Azim Premji University, Bangalore, 560100, India} and A.K. Sood$^{1}$}

\email{asood@iisc.ac.in}
 
\affiliation{$^{1}$Department of Physics, Indian Institute of Science, Bangalore 560012, India}
           
\date{\today}

\begin{abstract}
Mixed surfactant systems with strongly bound counterions show many interesting phases such as the random mesh phase consisting of a disordered array of defects (water-filled nano-pores in the bilayers). The present study addresses the non-equilibrium phase transition of the random mesh phase under shear to an ordered mesh phase with a high degree of coherence between nano-pores in three-dimension. In-situ small-angle synchrotron X-ray study under different shear stress conditions shows sharp Bragg peaks in the X-ray diffraction, successfully indexed to the rhombohedral lattice with R$\bar{3}$m space group symmetry. The ordered mesh phase shows isomorphic twinning and buckling at higher shear stress. Our experimental studies bring out rich non-equilibrium phase transitions in concentrated cationic surfactant systems with strongly bound counterions hitherto not well-explored and provide motivation for a quantitative understanding.
\end{abstract}

\maketitle


\section*{Introduction}

Mesh phases are liquid crystalline phases formed in ionic as well as non-ionic mixed surfactant systems, consisting of a 1D stack of bilayers with water-filled pores or curvature defects. Depending on the correlation of these pores, the phase can be distinguished as a random mesh phase (L$_\alpha^D$) if the pores exhibit a liquid-like ordering, or an ordered mesh phase if the pores are correlated across the bilayers by getting locked into a three-dimensional lattice. These can be further classified as tetragonal or rhombohedral (R$\bar{3}$m) mesh phase, depending on the structural symmetry \cite{kekicheff1989structure,krishnaswamy2005phase,gupta2013controlling}. As a simple manifestation, the mesh phase occurring at higher surfactant concentrations is structurally similar to the adjacent lamellar phase and topologically similar to the preceding hexagonal phase formed at lower surfactant concentrations. However, non-uniform mean curvature in the mesh phase is in sharp contrast with the hexagonal or lamellar phase \cite{cates1990statics,holmes2005bicontinuous}. Extensive studies on the equilibrium phase behavior of lyotropic surfactant systems have established that the spontaneous formation of the defects (pores) in the mesh phase and their long-range ordering across the stack of the bilayers require a balancing of the head group interactions and the chain flexibility, particularly through the addition of a third component to a binary surfactant-water mixture \cite{funari1992microscopy,leaver2001structural,krishnaswamy2005phase,ghosh2007structure,lucassen1981surface,manohar1986origin,kaler1992phase,yatcilla1996phase,blandamer2000titration}. Similarly, the perforated bilayers in lipid-water systems form intermediate phases due to the incorporation of proteins into the pores \cite{ rand1968x}. Random mesh phases and ordered mesh phases of different symmetry have been observed in different lyotropic mixed surfactant systems and their equilibrium phase behavior have been well studied both theoretically and experimentally \cite{hyde2003novel}. Interestingly, mesh phases are also observed in lipid-water systems, and are relevant in the context of membrane fusion \cite{yang2003rhombohedral}, cytolysis \cite{sakamoto2021direct}. Thus, a majority of the existing literature on mesh phases pertain to characterizing the structure of mesh phases in different ternary surfactant-water systems where these phases can be stabilized typically over a large range of surfactant concentration. 

When subjected to shear, pore free lyotropic lamellar phase  usually exhibits two kind of  non-equilibrium phase transitions (NEPT), one is rolling of lamellar into sphere like structures known as  multilamellar vesicles (MLV) or onions and another is  change in the orientation of the bilayers.  Lamellar to MLV transition was reported first time  by Diat et al. \cite{diat1993effect} and ever since has been reported in many surfactant  and polymeric systems and appears to be a common feature of lamellar phase \cite{bergenholtz1996formation,zipfel1999influence, koschoreck2009multilamellar,grobkopf2019shear}. The proposed mechanism for this transition is dilation strain in the bilayers under shear  which can  arise either due to suppression of undulations or dislocations in the plane of the bilayers \cite{zilman1999undulation}. Another transition namely, the change in the orientation occurs in a system where the dislocation in the bilayers can not follow shear flow and above critical shear rate the dilation around the dislocation destabilizes the orientation  from c-oriented state where bilayer normal is in gradient direction to  a-oriented  state where the bilayer normal is in vorticity direction \cite{zipfel1999influence}.  Also, the suppression of undulations in the bilayers can result in expulsion of solvent from the bilayers \cite{ramaswamy1992shear}. According to proposed models,  non-equilibrium transitions in the lamellar  have been attributed to the suppression of thermal undulations causing hydrodynamical instabilities that occur above a critical shear rate which is inversely proportional to cube of bilayer spacing and expected to occur in dilute liquid bilayer phases \cite{cates1989role,ramaswamy1992shear}. However lamellar to onion transition is observed in concentrated  bilayer forming phases well below critical shear rate predict by proposed theoretical models.  Albeit numerous attempts, the controlled parameter for shear induced lamellar to MLV transition remains unclear and most of these studies have been limited to bilayer systems which are stabilized by steric interactions and possess either edge or screw dislocation.

Though the kinetics of equilibrium phase transition from the mesh phase to the lamellar phase and role of temperature as well as counterions is well-understood \cite{raghunathan2012mesh,luzzati1960structure,kekicheff1991cylinders,fairhurst1997structure,kekicheff1989structure,holmes1998intermediate}, NEPT in mesh phases formed in concentrated mixed surfactant systems remain poorly studied despite the structural similarity between the unperforated lamellar phase and the mesh phase. In comparison, in concentrated mixed surfactant systems, above the Kraft temperature, the shear-induced reversible transition from pore-free bilayers to a crystalline phase is observed at intermediate shear rates originating due to the re-distribution of the counterions \cite{rathee2013reversible}. Similarly, shear-induced unbinding of counterions is observed in other bilayer forming or multilamellar vesicle phases \cite{mendes1997vesicle}. Also, an increase in the alignment of the randomly oriented crystallites of bilayers or cylinders over macroscopic dimensions in the plane of shear as well as a transition from random mesh phase to the onion phase have been observed under shear \cite{fairhurst1996shear,zipfel1999influence}. Notably, the transition to the onion phase is preceded by a change in orientation of the bilayers.

Here, we bring out a few unexplored aspects of the flow behavior of the mesh phases formed in cationic surfactant system with strongly bound counterions. Time-resolved Rheo-SAXS measurements allow us to follow the temporal evolution of the X-ray diffraction peaks corresponding to different lattice planes of the pores of the bilayers. A reorientation of the bilayers of the L$^{D}_{\alpha}$ phase is always observed under shear with the bilayers stacked parallel to the shear-plane ($\mathbf{V}$-$\nabla{\mathbf{V}}$ plane) which is identified as the a-orientation. Intriguingly, further shearing reveals two distinct structural transitions depending on the separation between bilayers of the L$^{D}_{\alpha}$ phase. At lower surfactant concentrations (larger bilayer separation), the a-oriented bilayers of the L$^{D}_{\alpha}$ phase transform to an onion phase under shear, whereas, at higher surfactant concentrations (smaller bilayer separation), a non-equilibrium phase transition (NEPT) from the L$^{D}_{\alpha}$ phase to the R$\bar{3}$m phase is observed. The NEPT from the L$^{D}_{\alpha}$ phase to the R$\bar{3}$m phase occurs through the onset of correlation of the nano-pores across the bilayers before they get locked into a three-dimensional rhombohedral lattice. This is followed by the shear-induced isomorphic twinning and buckling transition of the ordered mesh phase (R$\bar{3}$m) due to the hydrodynamic instability.

\section*{Experimental Details}

\subsection*{\label{materials}Materials}
Surfactants cetyltrimethylammonium bromide (CTAB) and cetylpyridinium chloride (CPC) from Sigma Aldrich were used without further purification. Sodium-3-hydroxy-2-naphthoate (SHN) was prepared by adding an equivalent amount of an aqueous solution of sodium hydroxide (NaOH) to an ethanol solution of 3-hydroxy-2-naphthoic acid (HNA). Ternary solutions of surfactant-SHN-water were prepared with deionized water (resistivity $\sim$ 18.2 M$\Omega$.cm) for the total weight fraction of surfactant + SHN ($\phi = \mathrm{(Surfactant+SHN)/(Surfactant+SHN+H_2O)}$) at the desired molar ratio of the two components ($\alpha = \frac{\mathrm{[SHN]}}{\mathrm{[Surfactant]}}$) \cite{krishnaswamy2005phase}. In equilibrium, the counterion SHN is known to be adsorbed at the micelle-water interface \cite{krishnaswamy2005phase,ghosh2007structure,gupta2013controlling}, thus decreasing the spontaneous curvature of the micellar aggregates formed by CTAB or CPC, transforming the cylindrical micelles to pore-free bilayers as $\alpha$ approaches 1. The equilibrium phase diagrams of these systems [see section-A in the Supplemental Material] have shown random and ordered mesh phases over a wide-range of surfactant concentration ($0.2 < \phi < 0.7$). In our present experiments, $\alpha$ was fixed at 1 for the CTAB-SHN-water system and 0.5 for the CPC-SHN-water system. The samples were well-sealed and left in an oven at $40^\circ$ C for 2 weeks to equilibrate \cite{gupta2013controlling}. As a guide, the samples appear inhomogeneous immediately after mixing and finally becomes homogeneous and brownish once equilibrated with no apparent inhomogeneities seen by the naked eyes. All the measurements were done with the freshly loaded sample.

\subsection*{\label{rheosals} Rheo-SALS experiments}
In-situ depolarized small-angle light scattering (SALS) measurements were performed along with rheology in a shear stress controlled rheometer (MCR 102, Anton Paar) fitted with a temperature controller [Fig. \ref{F1}(a)]. Parallel plate (PP) glass geometry of diameter 43 mm was used with 1 mm sample thickness. The laser beam (wavelength of 658 nm) was in the velocity gradient direction at a fixed position 15 mm from the plate center. SALS images on the white screen were recorded by an 8-bit color CCD camera (Lumenera, 0.75C, 1200$\times$980 pixels) fitted with a PENTAX TV Lens of focal length 12 mm.

\subsection*{\label{rheosaxs} Rheo-SAXS experiments}
\begin{figure*}
	\includegraphics[width=1.0\textwidth]{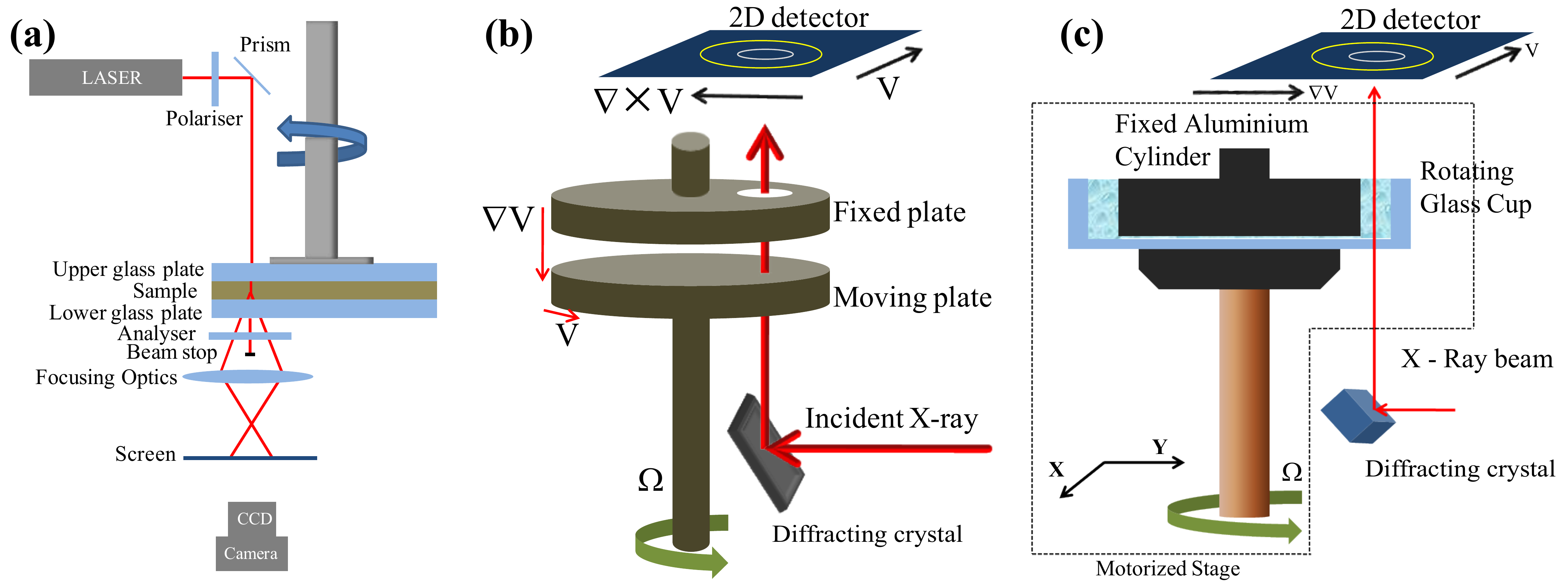}
	\caption{Schematic of the rheo-SALS setup with PP glass geometry (a). Schematics of the Rheo-SAXS setups with the PP vespel geometry (b), and with the Couette glass geometry (c). The rheo-SAXS setup with the Couette glass geometry was on a motorized stage to do the X-ray scan across the gap by varying the distance ($g_x$) of the X-ray from the inner static cylinder. Scattering wave vectors and different flow directions of the geometry are indicated.}
	\label{F1}
\end{figure*}

We have used time-resolved small-angle X-ray scattering (SAXS) setup coupled with Haake-Mars rheometer \cite{struth2011observation} at the P10 beamline of the PETRA III synchrotron [Fig. \ref{F1}(b),(c)]. The sample chamber was fitted with a Peltier based temperature controller and a humidity controller with a slow nitrogen flow. The synchrotron X-ray beam was deflected vertically and passed through the sample. The X-ray diffraction patterns were recorded on a Pilatus 300 K detector with a varying exposure time of 10-50 seconds. For the parallel plate (PP) vespel geometry (DuPont, diameter 35 mm), the sample thickness was chosen to be 1 mm due to the high absorbance, and the X-ray beam was in velocity gradient direction at a fixed position 13 mm from the plate center. The detector was placed parallel to the vorticity-plane ($\mathbf{V}$-$\nabla{\times\mathbf{V}}$ plane), at a distance 1000 mm from the sample. For the Couette glass geometry (with a metallic inner cylinder of diameter 29 mm and a outer glass cup of diameter 33 mm), the sample height was 5.5 mm, and the X-ray beam was parallel to the vorticity direction with the detector plane parallel to the shear-plane ($\mathbf{V}$-$\nabla{\mathbf{V}}$ plane). The position of the beam was varied in Couette's radial direction by slowly translating the motorized stage of the rheometer.

\section*{Results and Discussion}

Otherwise stated, here all the measurements were done at $30^{\circ}$ C (above the Kraft temperature of $25^{\circ}$ C) with the PP vespel geometry having sample thickness of 1 mm. Following the model in Ref. \cite{leaver2001structural}, we have estimated the size of the water-filled pores ($\sim \sqrt[3]{\mathrm{pore \:volume}}$) to be in the range from 3.7 nm to 4.6 nm [see section-B in the Supplemental Material] and hence justifying calling the pores as nano-pores.

\subsection*{Non-equilibrium phase transition from random mesh phase (L$^{D}_{\alpha}$) to onion phase} 

\begin{figure*}
	\includegraphics[width=1.0\textwidth]{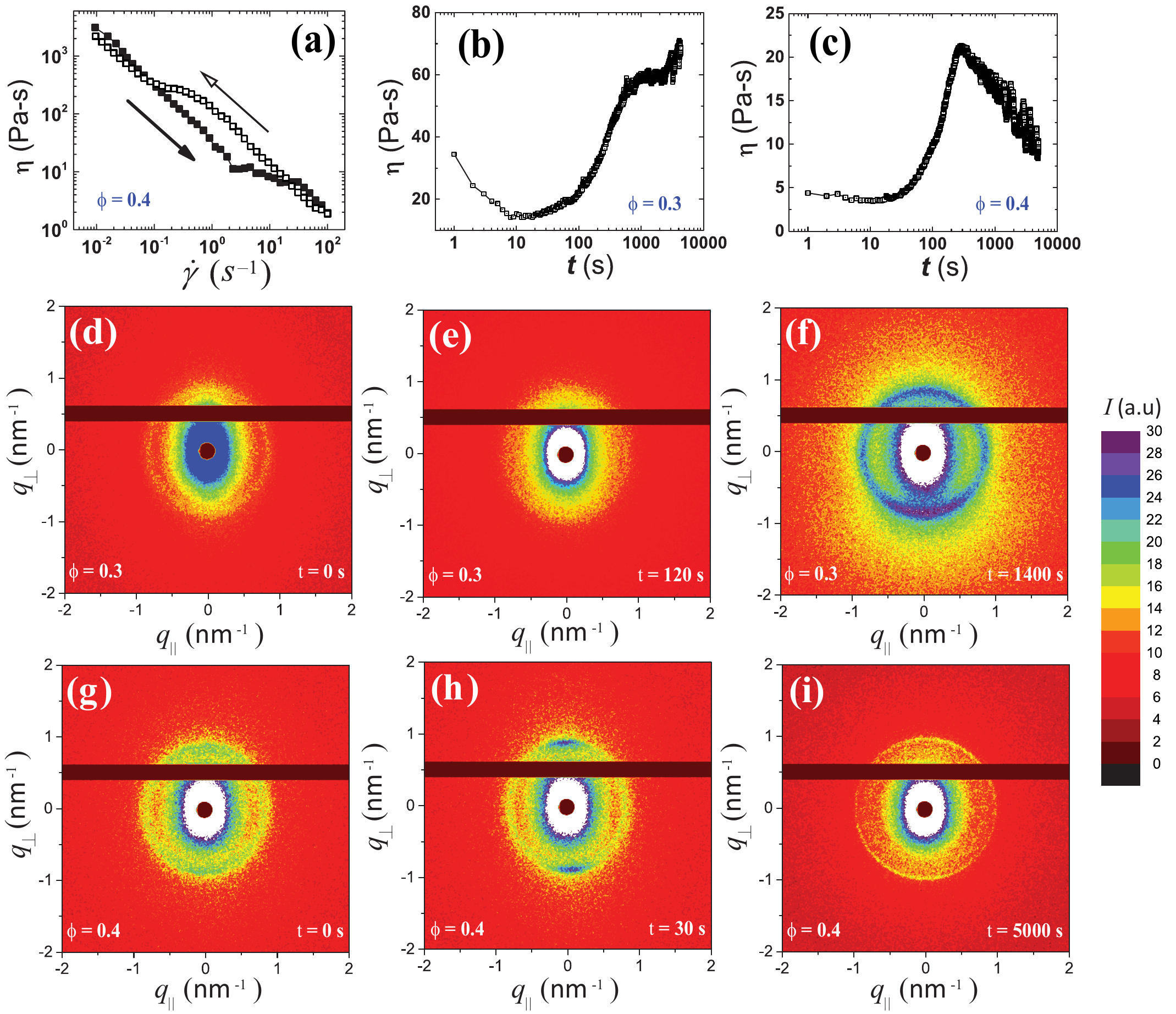}
	\caption{Shear rate controlled measurements of the random mesh phase (L$^{D}_{\alpha}$) formed in CTAB-SHN-water system ($\alpha = 1$) done in PP geometry. (a) Flow curve for $\phi = 0.40$; viscosity vs shear rate ($\eta$ vs $\dot{\gamma}$) with 30 s waiting time at each data point. Filled symbol and empty symbol represent increasing and decreasing $\dot{\gamma}$ respectively, as indicated by arrows. (b) Shear stress relaxation for $\phi = 0.30$; viscosity vs time ($\eta$ vs $t$) at $\dot{\gamma} = 10$ s$^{-1}$, and the observed X-ray diffraction pattern at $t = 0$ s (d), $t = 120$ s (e), $t = 1400$ s (f). (c) Similar shear stress relaxation for $\phi = 0.40$; $\eta$ vs $t$ at $\dot{\gamma} = 10$ s$^{-1}$, and the observed X-ray diffraction pattern at $t = 0$ s (g), $t = 30$ s (h), $t = 5000$ s (i). The intensity color scale (shown in the rightmost) is same for all the patterns in this figure.} 
	\label{F2}
\end{figure*}

We start with the CTAB-SHN-water system for $\phi <$ 0.5 and $\alpha = 1$ for which the flow curve is shown in Fig. \ref{F2}(a), recorded with the waiting time at each shear stress to be 30 sec. We point out that these flow curves are not steady-state flow curves as the system evolves with time under the application of shear stress. It can be seen that the random mesh phase (L$^{D}_{\alpha}$) exhibits a typical shear-thinning behavior where the viscosity ($\eta$) decreases with the shear rate or with the shear stress. At smaller shear rates, the lamellae sheets flow pass each other resulting in a sharp decrease in $\eta$ before showing an inflection point, followed by another shear-thinning region. In the reverse run, we observe only shear-thinning behavior suggesting that the observed structural changes are not shear-reversible. Since flow curve measurements are time-averaged over several seconds at a particular shear rate, time-resolved rheology reveals temporal evolution of viscosity at a constant shear rate. Figure \ref{F2}(b),(c) show the evolution of $\eta$ at a constant shear rate ($\dot{\gamma}$) of 10 s$^{-1}$, for two different volume fractions $\phi$ = 0.3 and 0.4 respectively at $\alpha$ = 1. For both values of $\phi$, $\eta$ shows a minimum at $t \sim 10$ s and then increases to a high value. For $\phi$ = 0.3, without shear, the X-ray diffraction pattern shows an isotropic Bragg ring having a very weak intensity due to the randomly oriented domains of lamellae having $d$-spacing $d_l = 9.60$ nm [Fig. \ref{F2}(d)]. Here, we do not observe the diffuse scattering peak due to the lack of in-plane correlation between nano-pores. Before $t = 1000$ s, the intensity of the isotropic Bragg ring becomes even lesser [Fig. \ref{F2}(e)], indicating a partially oriented lamellar phase with predominantly c-oriented bilayers. This is further evident from the low viscosity of the sheared sample at around 120 s compared to the starting phase, as the c-orientation of the bilayers (schematic is shown in Fig. \ref{F10}(B)) offers a lower resistance to shear. However, in the steady-state after $t = 1000$ s, a sharp isotropic Bragg ring is observed suggesting the formation of the onion phase under shear [Fig. \ref{F2}(f)]. For $\phi$ = 0.4, without shear, the observed unoriented lamellae [Fig. \ref{F2}(g)] has a $d$-spacing $d_l = 7.05$ nm and the nano-pores have in-plane correlation length $d_d = 7.68$ nm [Supplemental Material, Table S1]. For $t \geq 10$ s, a-orientation of the bilayers \cite{goulian1995shear} is observed (schematic is shown in Fig. \ref{F10}(A)) where the lamellar peak (the Bragg peak due to the lamellar periodicity) is sharp in the vorticity direction (in the $\nabla{\times\mathbf{V}}$ direction or $\mathbf{q_{\bot}}$ direction) and the diffuse scattering peak becomes sharp in the flow direction (in the $\mathbf{V}$ direction or $\mathbf{q_{\|}}$ direction), indicating bilayer planes stacked parallel to the $\mathbf{V}$-$\nabla{\mathbf{V}}$ plane (called the shear-plane) [Fig. \ref{F2}(h)]. Here we note that for $\phi$ = 0.4, the shear-alignment of the L$^{D}_{\alpha}$ phase was useful to distinguish the lamellar peak and the diffuse peak clearly as they are almost overlapping in the equilibrium diffraction pattern. The observed decrease in viscosity corresponding to the a-orientation indicates that the bilayers offer less resistance to shear in this configuration. After $t = 1000$ s, a sharp isotropic Bragg ring is observed suggesting the formation of the onion phase under shear [Fig. \ref{F2}(i)]. The observed induction time of about 10 s to 20 s in both the systems beyond which the viscosity increases is likely to be associated with the time to align a certain portion of the multi-domain L$^{D}_{\alpha}$ phase. The order of induction time is similar to other study \cite{escalante2000shear} where they show that the induction time decreases with increasing shear rate. The Fourier transform of the fluctuations in $\eta$ (Fig. \ref{F2}(b),(c)) gives a time scale of $\sim$ 12 sec (the plots are not shown) which is very close to the time period of one rotation of the plate confining the sample. In order to assign any meaning to these oscillations, much more studies in future are required to ascertain shear banding induced instabilities without the artefact of slip at the sample-plate interface. 

\begin{figure*}
	\includegraphics[width=1.0\textwidth]{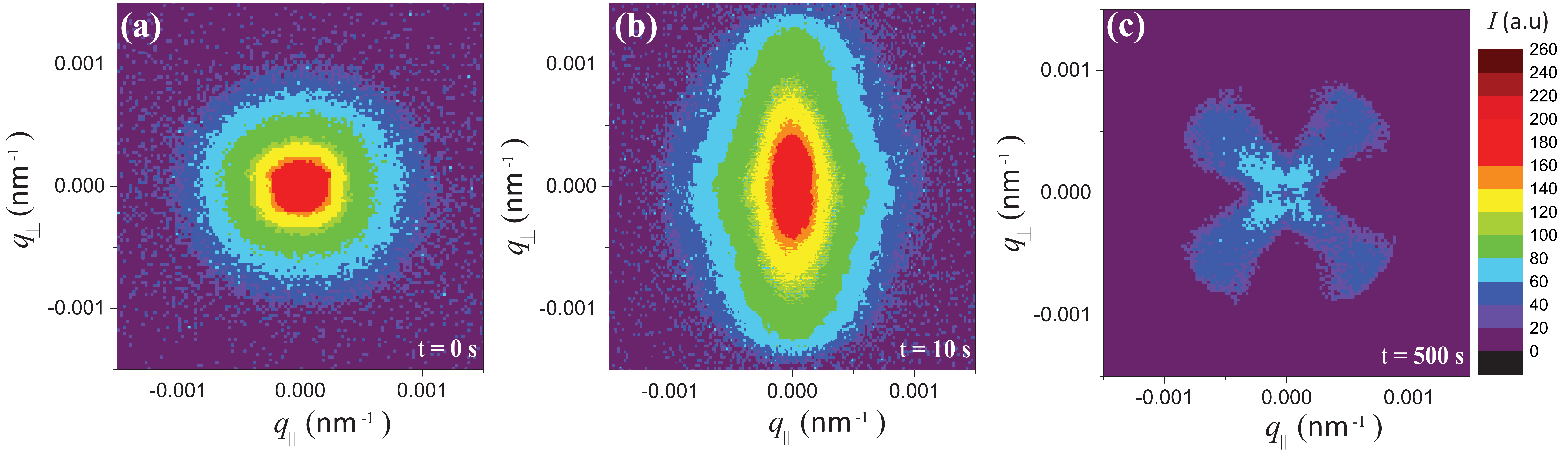}
	\caption{During the shear stress relaxation of L$^{D}_{\alpha}$ similar to Fig. \ref{F2}(c), small angle light scattering (SALS) patterns observed at $t = 0$ s (a), $t = 10$ s (b), $t = 500$ s (c) are shown ($\phi = 0.40$, $\alpha = 1$, $\dot{\gamma} = 10$ s$^{-1}$). The intensity color scale (shown in the rightmost) is same for all the patterns in this figure.} 
	\label{F3}
\end{figure*}

Further, we have used the in-situ small-angle light scattering (SALS) to confirm the NEPT from the L$^{D}_{\alpha}$ phase to the onion phase by observing a four-lobed clover-leaf pattern, a signature of the onion under the depolarized SALS \cite{zipfel1999influence}. The Rheo-SALS measurements were performed in the VH (polarizer $\bot$ analyzer) configuration. Figure \ref{F3} shows the evolution of Rheo-SALS patterns obtained during the shear stress relaxation measurement of the L$^{D}_{\alpha}$ phase for $\phi$ = 0.4 at $\dot{\gamma}$ = 10 s$^{-1}$. The time evolution of $\eta$ is the same as in Fig. \ref{F2}(c) (data not shown). At $t \sim 10$ s, the isotropic pattern [Fig. \ref{F3}(a)] changes to an anisotropic pattern [Fig. \ref{F3}(b)], oriented parallel to the $\mathbf{q_{\bot}}$ direction indicating the flow alignment of bilayers parallel to the shear-plane. After $t = 500$ s, the expected four-lobed clover-leaf pattern associated with the onion phase appears [Fig. \ref{F3}(c)]. The observed clover-leaf pattern in SALS arises from the optical anisotropy of the onions where two refractive indices that are parallel and perpendicular to the radial direction can be defined, and very similar to that obtained from spherulites of semi-crystalline polymers \cite{samuels1971small}. When the onion phase with the optic axis along the radial direction (normal to the bilayers) is placed between crossed polarizers, a four lobed pattern is obtained in SALS where maximum intensity is observed at $45^o$ and minimum at $0^o$, $90^o$. The increase in $\eta$ is due to the NEPT from well-separated lamellae to closely packed onions. After 1000 s, the observed fluctuations in the $\eta$ is possibly due to the onions' re-arrangements or variation in the size of these onions \cite{wunenburger2001oscillating}. Though the NEPT from L$^{D}_{\alpha}$ phase to onion phase under shear is a well-studied phenomenon \cite{zipfel1999influence}, the noteworthy aspect of the present study is that lamellar to onion transition is not limited to dilute lamellar phase or bilayers having defects in the form of dislocations but also observed in bilayers having defects in the form of solvent-filled pores. In our study, we find that L$^{D}_{\alpha}$ to onion transition occurs for bilayer separation greater than 7 nm. Before transition to the onion phase, the orientation of bilayers strongly depends on the pore spacing. In particular, we have observed a-orientation at $\phi$ = 0.3 and c-orientation at $\phi$ = 0.4. A schematic of this transition is shown in Fig. \ref{F10}(C).

\subsection*{Shear-induced 3D ordering of nano-pores in L$^{D}_{\alpha}$ phase} 

\begin{figure*}
	\includegraphics[width=1.0\textwidth]{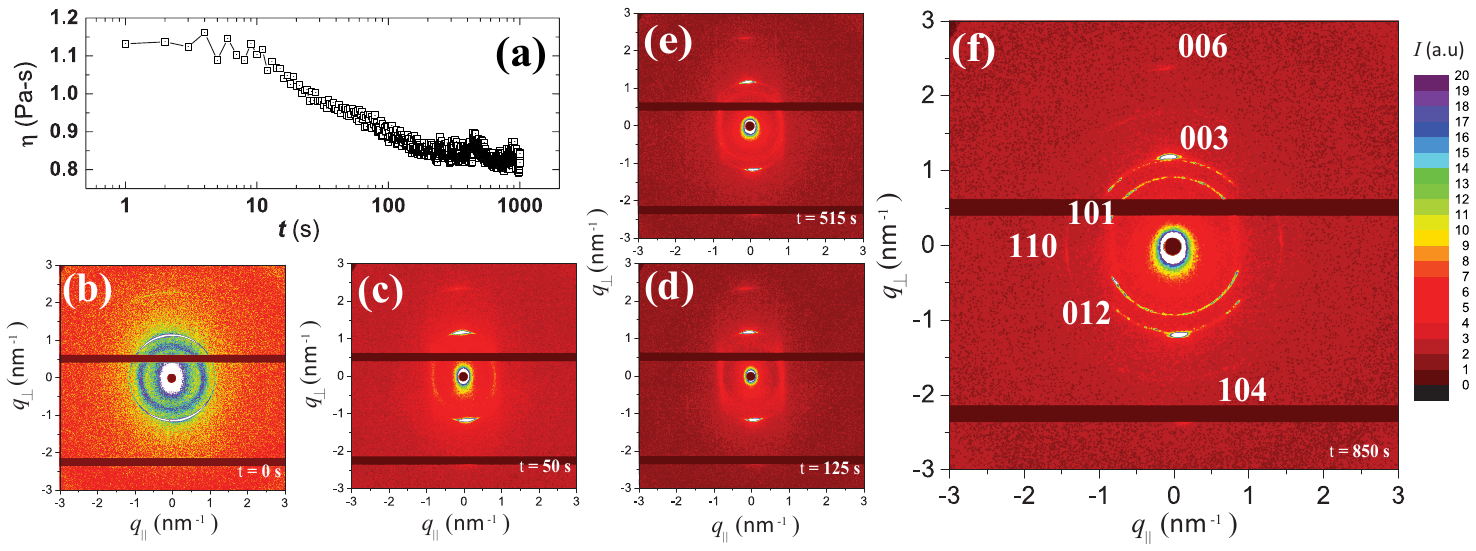}
	\caption{Shear stress relaxation of L$^{D}_{\alpha}$ for $\phi = 0.50$ (CTAB-SHN-water; $\alpha = 1$) at $\dot{\gamma} = 50$ s$^{-1}$ done in PP geometry. (a) $\eta$ vs $t$, and the observed X-ray diffraction patterns at $t = 0$ s (b), $t = 50$ s (c), $t = 125$ s (d), $t = 515$ s (e), $t = 850$ s (f) are shown. R$\bar{3}$m lattice planes are marked near the observed Bragg rings in (f). The intensity color scale (shown in the rightmost) is same for all the patterns in this figure.} 
	\label{F4}
\end{figure*}

The concentrated L$^{D}_{\alpha}$ phase formed in the CTAB-SHN-water system at $\phi$ = 0.5 and $\alpha$ = 1, demonstrates a NEPT from the L$^{D}_{\alpha}$ phase to the R$\bar{3}$m phase under shear. On applying a constant shear rate $\dot{\gamma} = 50$ s$^{-1}$, $\eta$ of the L$^{D}_{\alpha}$ phase decreases with time and it reaches the steady-state after $t \sim 200$ s [Fig. \ref{F4}(a)]. Figure \ref{F4}(b--f) show the temporal evolution of the X-ray diffraction pattern. In the quiescent state, the X-ray diffraction pattern [Fig. \ref{F4}(b)] reveals two isotropic Bragg rings with their $q$ ratio 1:2 (characteristic of an unaligned lamellar phase), with $d_l = 5.49$ nm [Supplemental Material, Table S1]. As expected in a random mesh phase, the diffuse isotropic ring observed at a smaller angle confirms the existence of liquid-like correlated nano-pores (water-filled) in the plane of the bilayer with an average in-plane correlation length $d_d = 7.67$ nm. The nearly isotropic rings corresponding to the lamellar periodicity of bilayers and the liquid-like correlation of the in-plane nano-pores, evolve to the well-aligned anisotropic diffraction pattern at $t \sim 50$ s, revealing the a-oriented state of the bilayers [Fig. \ref{F4}(c)]. The occurrence of the diffuse scattering peaks as arcs azimuthally centered at $\mathbf{q_{\|}} \sim$ 0 suggests an absence of the trans-bilayer correlation of the nano-pores. Notably, at $t \sim 515$ s, the azimuthal intensity profile of the diffuse scattering peak shows a splitting away from $\mathbf{q_{\|}}$ = 0 [Fig. \ref{F4}(e)], suggesting the onset of long-range correlation of the nano-pores across the bilayers favoring the formation of a 3D ordered structure. At $t \sim 850$ s [Fig. \ref{F4}(f)] a few more peaks appear in the X-ray diffraction pattern, with no further change upon shearing up to 1000 s, indicating a steady-state. The temporal evolution of the SAXS-diffractogram during this transition is presented in the section-C in the Supplemental Material (Fig. S3). All the partially oriented Bragg reflections of the diffraction pattern shown in Fig. \ref{F4}(f) could be indexed to a rhombohedral lattice with the R$\bar{3}$m space group with lattice parameters $a = 8.68$ nm and $c = 15.93$ nm, coexisting with the L$^{D}_{\alpha}$ phase [Supplemental Material, Table S3]. The lamellar reflection of the L$^{D}_{\alpha}$ phase overlaps with the (003) reflection of R$\bar{3}$m phase (lamellar periodicity $\sim c/3 = 5.31$ nm) and its coexistence with the R$\bar{3}$m phase in the final steady-state is inferred from the diffuse arcs azimuthally centered at $\mathbf{q_{\|}} \sim$ 0. With both the phases in the steady-state, shear-banding situation is possible where both the phases experience different shear rates.

We have chosen the following R$\bar{3}$m-indexing scheme where the first reflection overlapping with the diffuse scattering peak from the nano-pores is indexed as (101) and the third reflection which lies on the $\mathbf{q_{\bot}}$ axis as the (003) reflection. As seen from Table S3 in the Supplemental Material, the second reflection appearing as the shoulder of the (003) reflection is indexed as the (012) reflection. Moreover, the (110) reflection corresponding to scattering from the in-plane nano-pores of the bilayers lie parallel to $\mathbf{q_{\|}}$, further confirming the robustness of our indexation to the diffraction peaks of the R$\bar{3}$m structure. The structure of the shear-induced R$\bar{3}$m phase could be modeled as an ordered mesh phase with ABC stacking of 3-connected rods \cite{ghosh2007structure}. Moreover, the lattice parameters obtained for the shear-induced R$\bar{3}$m phase are similar to those obtained for the equilibrium R$\bar{3}$m phase at $\phi = 0.53$. Using the lattice parameters ($a$, $c$) and the above discussed 3-connected rod model, the estimated value of the micellar radius ($r_m$) is 2.10 nm, consistent with the equilibrium value reported in Ref. \cite{ghosh2007structure}. We propose that during this NEPT the decrease in viscosity $\eta$ is due to the persistence of a-orientation of the bilayers that offers low resistance to shear, as well as is due to the three-dimensional ordering of the nano-pores (schematic is shown in Fig. \ref{F10}(D)) that provides the passage for the liquid to pass through the connected pores. To quantify the reversibility of the L$^{D}_{\alpha}$ to R$\bar{3}$m transition, we stopped the shear after 1000 s and follow the SAXS diffraction pattern with time. Since the (003) reflection corresponds to both the L$^{D}_{\alpha}$ and R$\bar{3}$m phases, we follow (101) and (110) Bragg peaks of the shear-induced R$\bar{3}$m whose intensities decrease with time [see section-C in the Supplemental Material, Fig. S4], confirming reversibility of the transition. 

To ascertain that the NEPT from L$^{D}_{\alpha}$ phase to R$\bar{3}$m phase is generic, at least to other cationic surfactant system with strongly bound counterions, we have also studied the concentrated L$^{D}_{\alpha}$ phase (with $d_l = 5.13$ nm and $d_d = 6.50$ nm) formed in CPC-SHN-water system ($\phi$ = 0.55 and $\alpha$ = 0.5) \cite{gupta2013controlling}. In this case also, we have observed a similar NEPT from the L$^{D}_{\alpha}$ phase to the R$\bar{3}$m phase [see section-D in the Supplemental Material]. The viscosity $\eta$ decreases with time but reaches the steady-state after $t \sim 20$ s [section-D in the Supplemental Material, Fig. S5(a)], in a comparatively short period of time (compared to the CTAB-SHN-water system shown in Fig. \ref{F4}). Under shear, the L$^{D}_{\alpha}$ phase goes to the coexistence of two R$\bar{3}$m phases with lattice parameters $a1 = 8.39$ nm, $c1 = 14.79$ nm and $a2 = 8.27$ nm, $c2 = 14.28$ nm [Supplemental Material, Table S2]. After stopping the shear, the temporal evolution of the SAXS-diffractogram shows a complete reversibility from the shear-induced R$\bar{3}$m phase to the L$^{D}_{\alpha}$ phase within 10 seconds [section-D in the Supplemental Material, Fig. S6]. From our observations on these two systems, we suggest that the shear-induced ordering of the water-filled nano-pores is expected in systems where both the random mesh phase and the ordered mesh phase are present adjacent to each other in the equilibrium phase diagram. However, the control parameters of the transition will depend on the d-spacing, stabilizing forces, water pore size, etc.

\subsection*{Mechanism of the shear-induced R$\bar{3}$m ordering of the nano-pores of L$^{D}_{\alpha}$}

The shear-induced R$\bar{3}$m phase obtained in two different systems on applying a constant shear rate indicates that this NEPT is a general feature of concentrated random mesh phases. The distinguishing feature governing the kinetics of the NEPT from L$^{D}_{\alpha}$ phase to R$\bar{3}$m phase at a constant shear rate, that may be identified in both these systems from the time-resolved Rheo-SAXS measurements, is the a-orientation of the lamellae prior to the appearance of the sharp (101) reflection corresponding to the 3D ordering of nano-pores. In the presence of thermal undulations, the a-orientation of bilayers in the lamellar phase is usually preferred at high shear rates since the suppression of thermal undulations under shear is lower in the a-orientation in comparison with the c-oriented state (where the bilayer planes are stacked parallel to the $\mathbf{V}$-$\nabla{\times\mathbf{V}}$ plane, called the vorticity-plane) \cite{bruinsma1992shear}. 

We propose that the locking of the nano-pores into a 3D lattice occurs when the in-plane correlation length of the nano-pores ($d_d$) is larger than the bilayer periodicity ($d_l$). A comparison can be drawn here with respect to the equilibrium phase behavior of the system (CTAB-SHN-water system, $\alpha$ = 1), where the L$^{D}_{\alpha}$ phase to the R$\bar{3}$m phase transition is observed with decreasing water content (for $\phi >$ 0.5). The in-plane periodicity ratio ($\Gamma$) is estimated from the ratio of $d_d$ (or lattice parameter a in case of R$\bar{3}$m) to the bilayer separation $d_l$ for the L$^{D}_{\alpha}$ phase (or lattice parameter $c/3$ in case of R$\bar{3}$m). We find that when the surfactant volume fraction $\phi$ increases from 0.5 to 0.53, $\Gamma$ increases from a value of 1.2 in the L$^{D}_{\alpha}$ phase to 1.4 in the R$\bar{3}$m phase. A crucial aspect favoring our argument would be an increase in $d_d$ with shear observed from the diffuse scattering peak positions. However, domains of L$^{D}_{\alpha}$ with an increased $d_d$ is not observed in our experiments possibly because the strong flow imposed on the sample at a high shear rate, smears out the diffraction pattern from these domains, giving rise to a broad and diffuse scattering peak. Nevertheless, it should be noted that an increase in pore size at a constant water content ($\phi$) implies that water from the inter-bilayer region will enter into the nano-pores, thus decreasing the bilayer separation. Hence the observed decrease in the lamellar periodicity by $\sim$ 3 \AA under shear reinforces the proposed shear-induced increase in average pore size. Further, the consequent increase in $\Gamma$ from 1.3 to 1.6 for the shear-induced R$\bar{3}$m phase is consistent with the increase in $\Gamma$ observed for the L$^{D}_{\alpha}$ phase to the R$\bar{3}$m phase transition on decreasing water content in equilibrium \cite{ghosh2007structure}.

We note that the equilibrium phase behavior of the CTAB-SHN-water system \cite{krishnaswamy2005phase} indicates that for higher molar ratio of SHN to CTAB ($\alpha > 1$), the R$\bar{3}$m mesh phase can occur at a lower surfactant concentration. Here, under shear, a rearrangement of the organic salt/counterions on the bilayers can increase $\alpha$ locally, driving the transition to the R$\bar{3}$m mesh phase at lower surfactant concentrations. Similar shear-induced phase transitions have been reported in cationic-anionic mixed surfactant systems \cite{rathee2013reversible} where transition from an isotropic to crystalline, as well as lamellar to crystalline phases were observed. It is noteworthy that the NEPT from the L$^{D}_{\alpha}$ phase to the R$\bar{3}$m phase is absent at lower surfactant volume fractions for $\phi <$ 0.5, having $\alpha = 1$. A robust conclusion that emerges from our studies on the concentrated random mesh phase is that a shear-induced ordering of the membrane nano-pores occludes the formation of onion phases.

\subsection*{Plastic deformation of R$\bar{3}$m phase during flow}

\begin{figure*}
	\centering
	\includegraphics[width=1.0\textwidth]{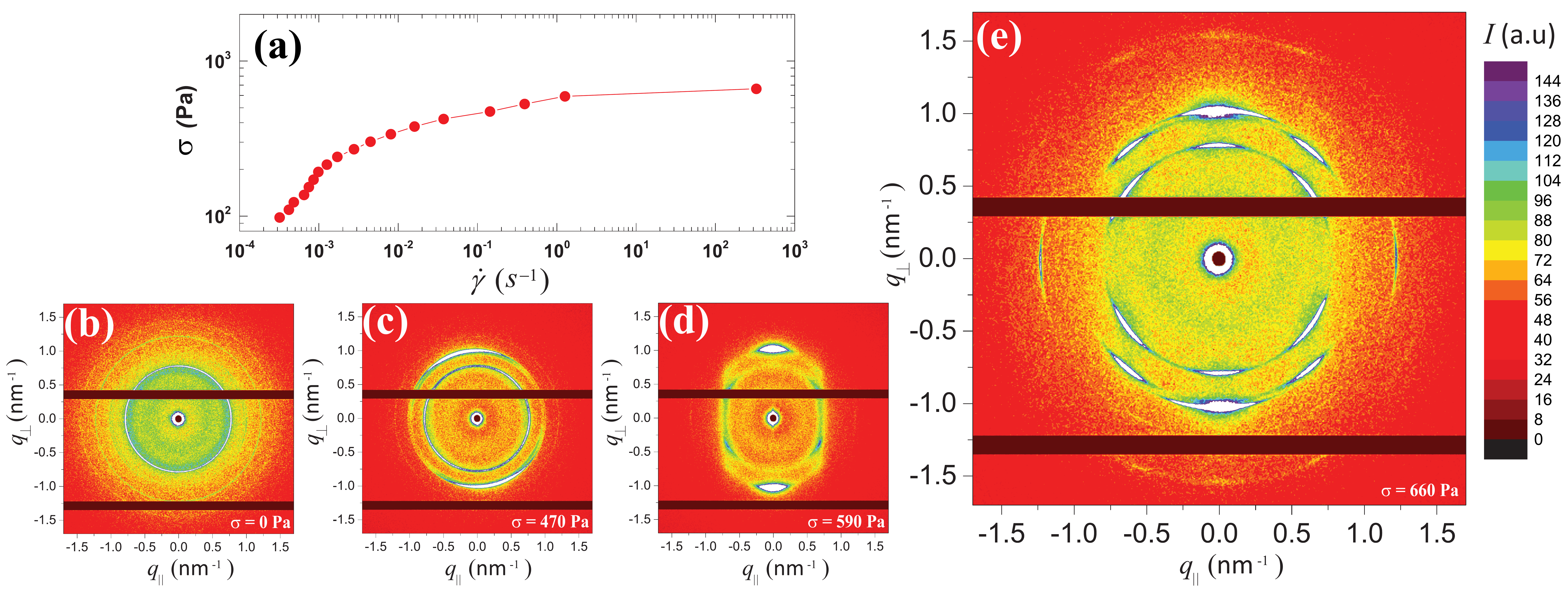}
	\caption{Shear stress controlled flow curve of the rhombohedral mesh phase (R$\bar{3}$m) formed in CTAB-SHN-water system ($\alpha = 1$, $\phi = 0.53$) done in PP geometry. (a) Shear stress vs shear rate ($\sigma$ vs $\dot{\gamma}$) plot with 200 s waiting time at each data point. Observed X-ray diffraction pattern for $\sigma = 0$ Pa (b), $\sigma = 470$ Pa (c), $\sigma = 590$ Pa (d), $\sigma = 660$ Pa (e) are shown. The intensity color scale (shown in the rightmost) is same for all the patterns in this figure.}
	\label{F5}
\end{figure*}

We will now present the effects of shear flow on the randomly oriented crystallites of the R$\bar{3}$m phase for different values of $\phi$ [see section-E in the Supplemental Material for the SALS and the SAXS measurements with the equilibrium R$\bar{3}$m phase]. Figure \ref{F5}(a) shows the shear stress controlled flow curve of the R$\bar{3}$m phase (CTAB-SHN-water, $\phi = 0.53, \alpha = 1$) with a stepwise increment in shear stress with 200 s waiting time at each data point. The waiting time of 200 s is not enough to reach strain of order 1 and hence the inflection point at strain rate of $\sim$ 0.001 s$^{-1}$ may not correspond to a physical effect. The flow curve reported at small shear rate (below 5 $\times$ 10$^{-3}$ s$^{-1}$) is only qualitative.

The diffraction pattern of the randomly aligned sample at $\sigma = 0$ Pa [Fig. \ref{F5}(b)] progressively transforms to the perfectly aligned sample (a-oriented state) at $\sigma = 590$ Pa with $\dot{\gamma} \sim 1$ s$^{-1}$ [Fig. \ref{F5}(c),(d)]. The X-ray diffraction pattern of the perfectly aligned phase [Fig. \ref{F5}(d)] has four diffuse arcs in (101), (012) Bragg rings and has two concentrated arcs in the (003) Bragg ring, consistent with the rotational symmetry of reciprocal lattice points of R$\bar{3}$m \cite{yang2003rhombohedral}. At $\sigma = 660$ Pa (corresponding $\dot{\gamma} \sim 100$ s$^{-1}$), six arcs are observed in (101), (012) Bragg rings, respectively, which we propose are due to buckling \cite{silmore2020buckling} with two different states of orientation i.e. presence of two directors [Fig. \ref{F5}(e)] (schematic is shown in Fig. \ref{F10}(E)), consistent with the observed star-like pattern in the Rheo-SALS measurement [see section-F in the Supplemental Material]. In the Rheo-SALS, the pattern is wide in two different directions (nearly orthogonal) which can be interpreted that the system has two preferred directions of bilayer orientations during flow. The lattice parameters of R$\bar{3}$m remain the same before and after the buckling [Supplemental Material, Table S4]. For $\sigma > 700$ Pa, an extreme shear-thinning is observed.

\begin{figure*}
	\centering
	\includegraphics[width=1.0\textwidth]{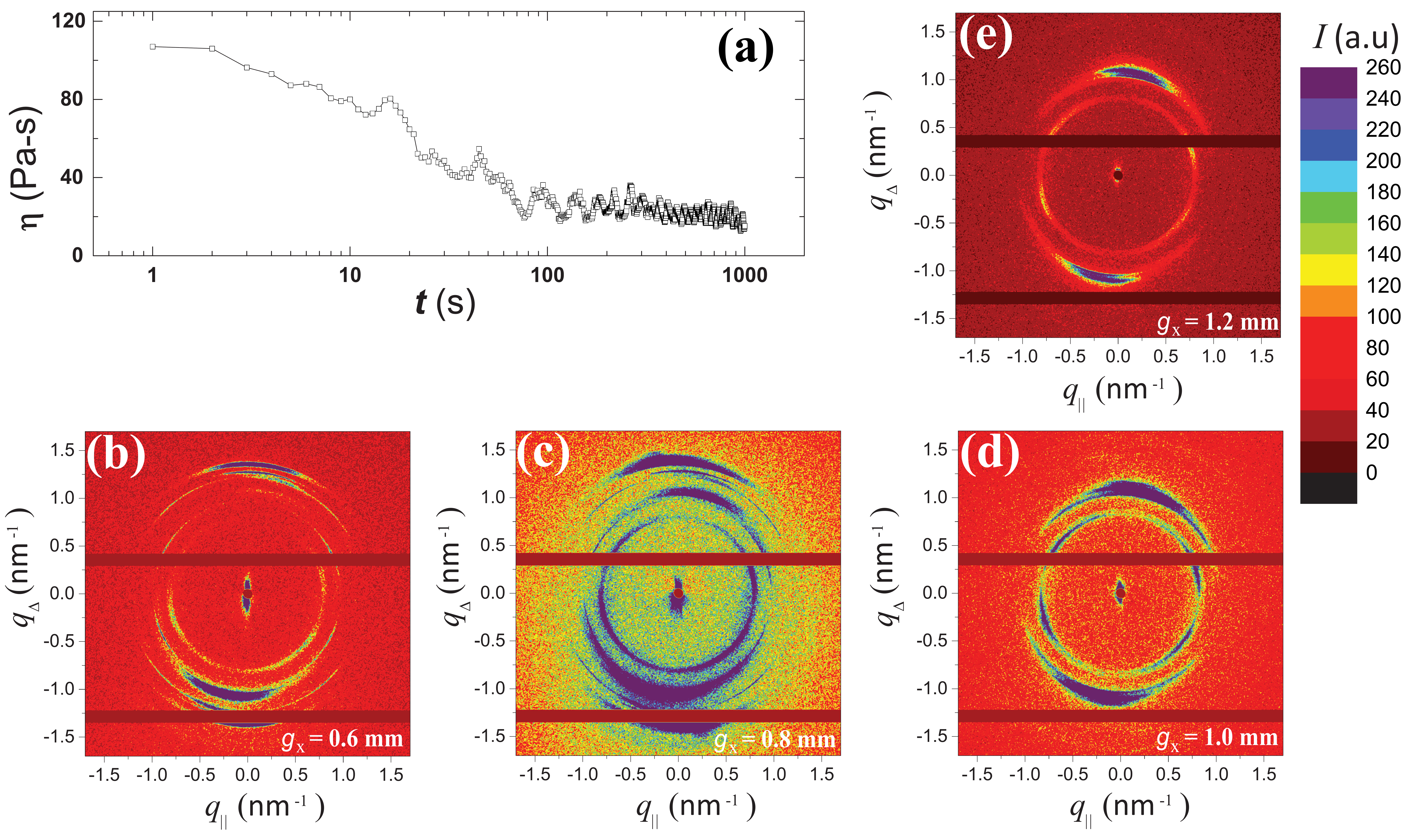}
	\caption{Shear stress relaxation of R$\bar{3}$m (CTAB-SHN-water system, $\alpha = 1$, $\phi = 0.53$) at $\dot{\gamma} = 1$ s$^{-1}$ done in Couette geometry. (a) $\eta$ vs $t$, and the X-ray diffraction patterns observed during 700 s $\leq t \leq 850$ s, with the incident X-ray beam at different distances ($g_x$) from the inner static cylinder; (b) $g_x = 0.6$ mm, (c) $g_x = 0.8$ mm, (d) $g_x = 1.0$ mm, (e) $g_x = 1.2$ mm. The intensity color scale (shown in the rightmost) is same for all the patterns in this figure.}
	\label{F6}
\end{figure*}

To probe this transition at different positions in the shear gradient direction in PP geometry, the X-ray has to pass through the sample at an angle to the vorticity direction that is not feasible in the present scattering geometry. To overcome this difficulty, we have used the Couette geometry as discussed below. The X-ray beam was translated in the gradient direction to place the beam at different positions ($g_x$) in the Couette gap. In the PP geometry, the structural transition occurs at $\dot{\gamma} \sim 1$ s$^{-1}$, thus here we have followed the temporal evolution of the X-ray diffraction patterns during the shear stress relaxation at $\dot{\gamma} = 1$ s$^{-1}$ [Fig. \ref{F6}]. The viscosity shows a monotonic decay up to $\sim 100$ s and then fluctuates about 20 Pa-s in the steady-state. The X-ray diffraction patterns were recorded after 700 s. Remarkably, an unexpected rich sequence of orientational transition accompanies the coexistence of two R$\bar{3}$m phases as discussed below. The equilibrium sharp isotropic (003) Bragg ring becomes sharp in the velocity gradient direction (the $\nabla{\mathbf{V}}$ direction or $\mathbf{q_{\Delta}}$ direction), and the aligned diffraction pattern is called the c-oriented state, indicating the correlated bilayer planes are stacked parallel to the vorticity-plane. The X-ray diffraction patterns for four different $g_x$ are showing the coexistence of two R$\bar{3}$m phases with different lattice parameters [Supplemental Material, Table S5]. The relative orientation of these two R$\bar{3}$m phases changes with $g_x$ as the incident X-ray beam is taken from the stator towards the rotor. Our measurements with the R$\bar{3}$m phase in both the shearing-geometries suggest that both “a” as well “c”-oriented bilayers coexist. However, with the shearing-geometries used, we cannot probe these two orientations simultaneously. In particular, here, the a-orientation (c-orientation) will not be observed in plate-plate (Couette cell) geometry.

\begin{figure*}
	\centering
	\includegraphics[width=1.0\textwidth]{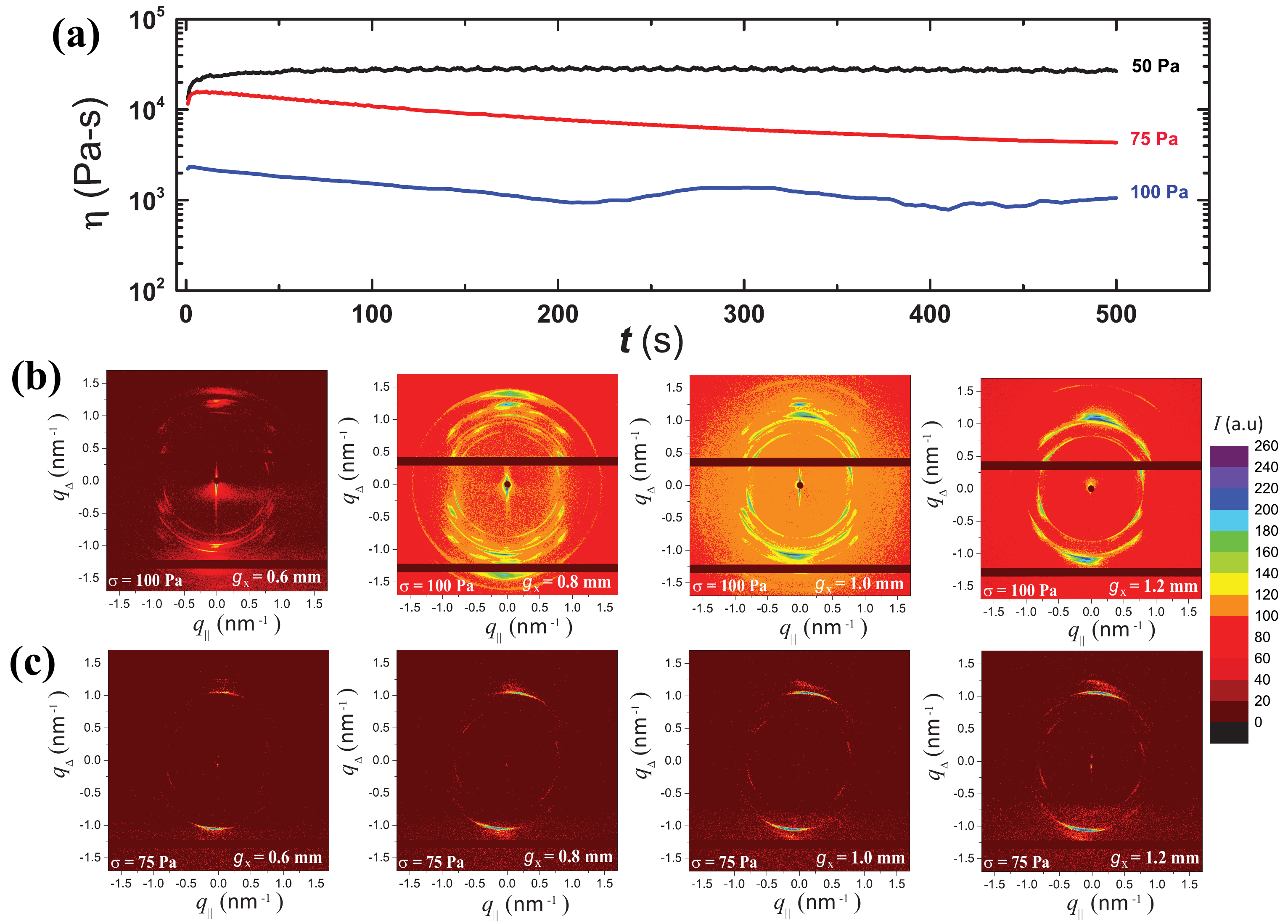}
	\caption{Shear rate relaxation of R$\bar{3}$m (CTAB-SHN-water system, $\alpha = 1$, $\phi = 0.53$) done in Couette geometry. (a) $\eta$ vs $t$ at different applied $\sigma$. X-ray diffraction patterns for different $g_x$ are shown; row (b) for $\sigma = 100$ Pa, and row (c) for $\sigma = 75$ Pa. For each measurement a fresh sample was loaded and the SAXS was done during 300 s $\leq t \leq 450$ s. The intensity color scale (shown in the rightmost) is same for all the patterns in this figure.}
	\label{F7}
\end{figure*}

We have also performed the shear rate relaxation measurements on an unaligned R$\bar{3}$m phase in Couette geometry [Fig. \ref{F7}]. For $\sigma < 100$ Pa, the system flows with a partially c-oriented R$\bar{3}$m throughout the gap between shearing cylinders [Fig. \ref{F7}(c)]. On applying high shear stress ($\sigma = 100$ Pa), the diffraction pattern shows several Bragg rings [Fig. \ref{F7}(b)]. Coexistence of two R$\bar{3}$m phases with different lattice parameters [Supplemental Material, Table S6] is observed in all the diffraction patterns for all $g_x$, except for $g_x = 1.2$ mm (here only the starting R$\bar{3}$m phase is observed). The relative orientation of these two R$\bar{3}$m changes with $g_x$ as one goes from stator to the rotor. The presence of six or eight arcs in (101), (012) Bragg rings near the inner static cylinder [Fig. \ref{F7}(b)] suggests buckling of R$\bar{3}$m with two different states of the orientation of nano-crystallites with the same lattice parameters. The lattice parameters $a$, $c$ of the two R$\bar{3}$m phases do not show any significant change in the shear gradient direction [see section-G in the Supplemental Material], suggesting sample-uniformity during shear. The isomorphic twinning of the R$\bar{3}$m phase has substantial variation in the relative orientation of the directors as one goes from stator to the rotor. This suggests that the buckling angle can vary in velocity gradient direction, and the crystallites corresponding to different directors may have different lattice parameters.

Now we correlate the modulation in the X-ray diffraction to the domains' orientation deep inside the sample under shear (for example part (d) and (e) of Fig. \ref{F5}). At a moderate shear rate, the R$\bar{3}$m phase has an a-aligned state with all layers facing towards vorticity direction. The X-ray diffracting in a perpendicular direction to the layer normal will show two strong arcs in the (003) Bragg ring, and four arcs in (012), (101) Bragg rings due to lattice symmetry of R$\bar{3}$m. At a high shear rate, few layers still retain the a-oriented state whereas others reorient themselves at a certain angle ($\sim 45^\circ$) with the vorticity direction, giving rise to more than four arcs in (012), (101) Bragg rings. The absence of many arcs in the (003) Bragg ring can be due to the fact that the (003) diffraction has only two-fold symmetry and the reoriented layers can easily miss it due to the finite angle with the shear gradient direction. This scenario is depicted in Fig. \ref{F10}(D). Note, similar model has been used to explain the small-angle scattering data from the Kraton-type block copolymers during tensile deformation where four arcs were observed in place of two in the Bragg ring corresponding to the spacing between cylinders \cite{hamley2001structure}. 

\begin{figure*}
	\includegraphics[width=1.0\textwidth]{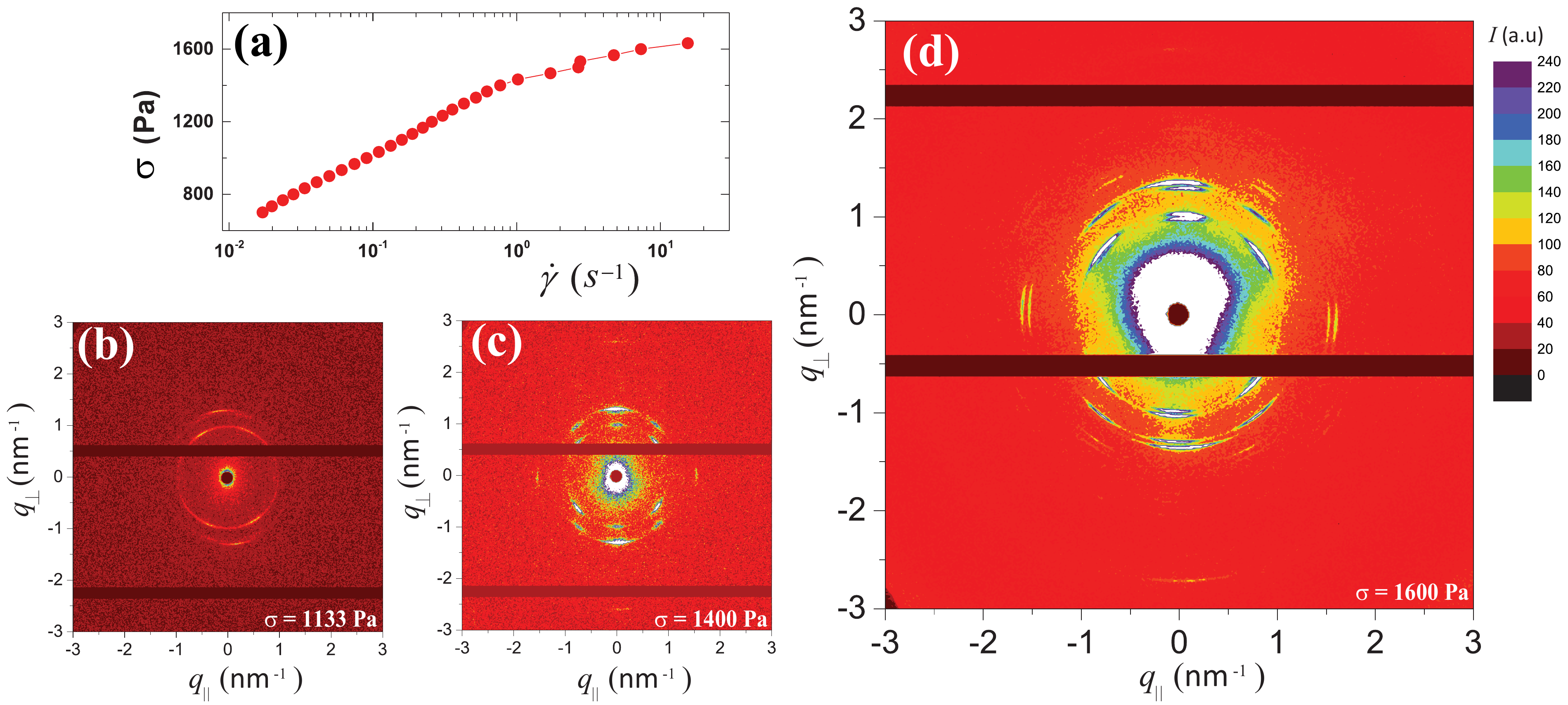}
	\caption{Shear stress controlled flow curve of R$\bar{3}$m (CTAB-SHN-water, $\alpha$ = 1, $\phi$ = 0.6) done in PP geometry. (a) $\sigma$ vs $\dot{\gamma}$, and the observed X-ray diffraction patterns for $\sigma$ = 1133 Pa (b), $\sigma$ = 1400 Pa (c), $\sigma$ = 1600 Pa (d) are shown. The intensity color scale (shown in the rightmost) is same for all the patterns in this figure.} 
	\label{F8}
\end{figure*}

\begin{figure*}
	\includegraphics[width=1.0\textwidth]{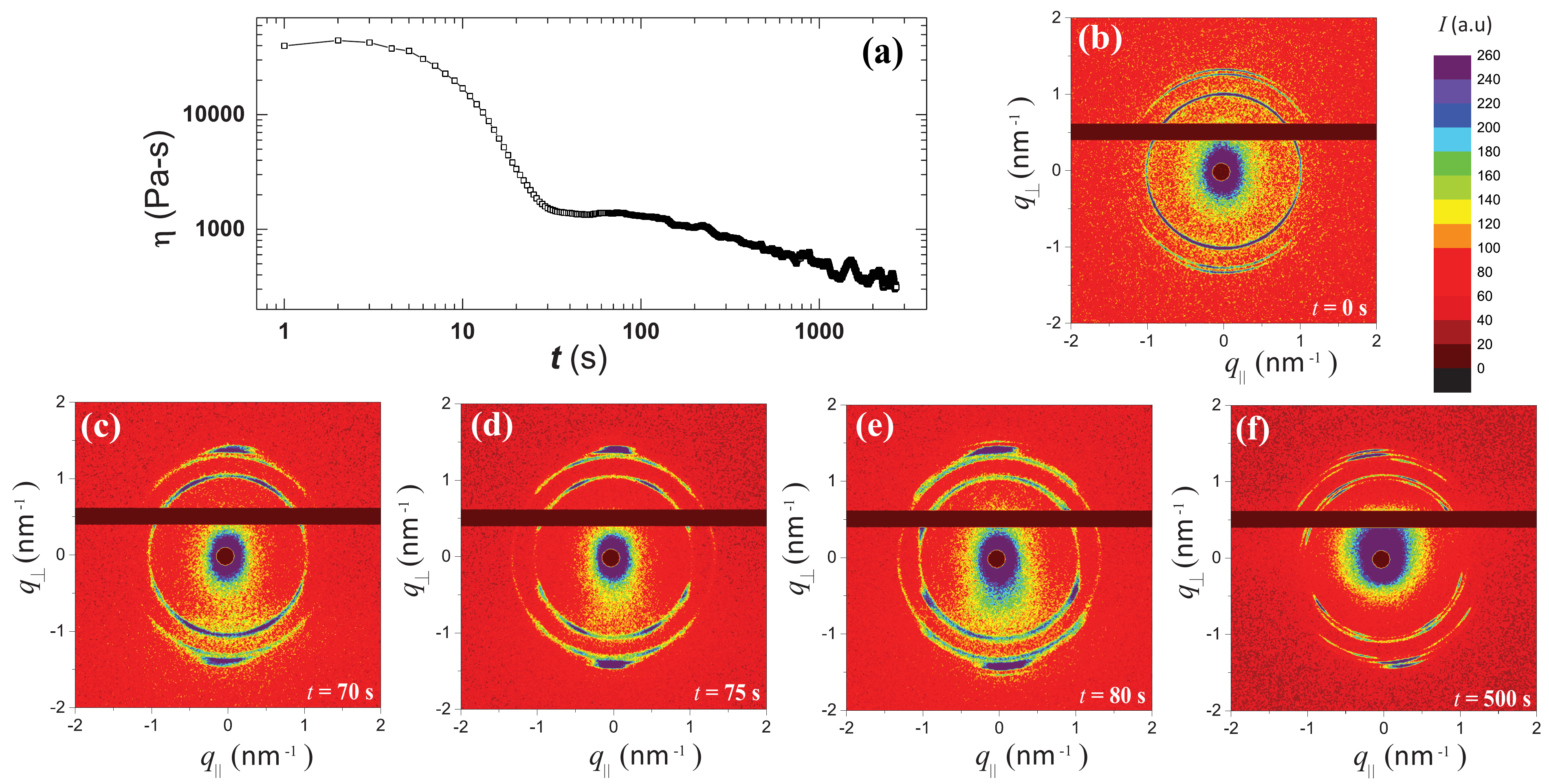}
	\caption{Shear stress relaxation of R$\bar{3}$m (CTAB-SHN-water system, $\alpha = 1$, $\phi = 0.60$) at $\dot{\gamma}$ = 1 s$^{-1}$ done in PP geometry. (a) $\eta$ vs $t$, and the observed X-ray diffraction patterns at $t = 0$ s (b), $t = 70$ s (c), $t = 75$ s (d), $t = 80$ s (e), $t = 500$ s (f) are shown. The intensity color scale (shown in the right corner) is same for all the patterns in this figure.}
	\label{F9}
\end{figure*}

\begin{figure*}
	\includegraphics[width=1.0\textwidth]{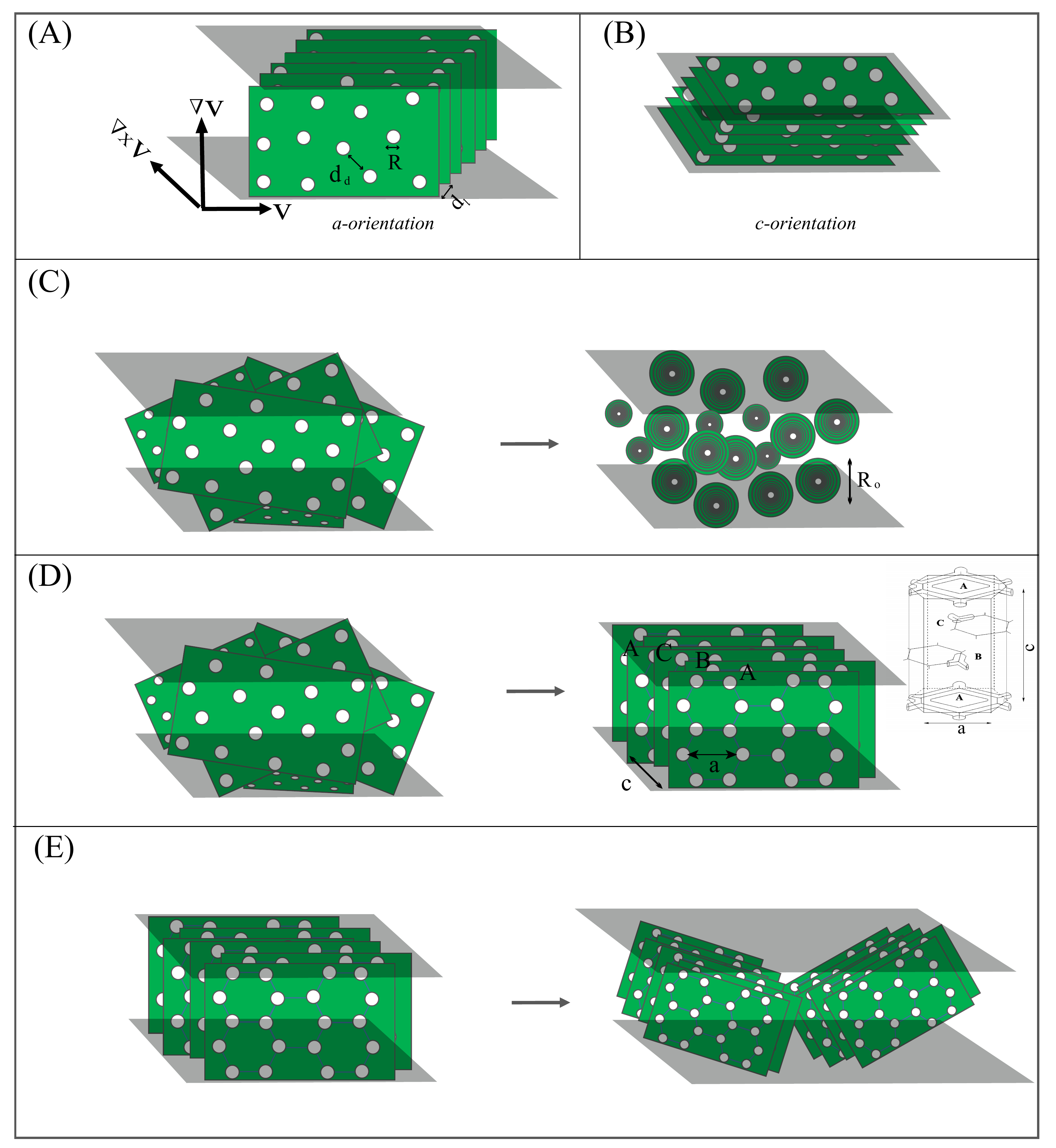}
	\caption{Schematics showing ‘a’–orientation of random mesh phase (A), ‘c’–orientation of random mesh phase (B), random mesh phase to onion phase transition (C), random mesh phase to ordered mesh phase transition (D), buckling transition of ordered mesh phase (E). An unit cell of the ordered mesh phase is shown in (D).}
	\label{F10}
\end{figure*}

The concentrated R$\bar{3}$m phase (CTAB-SHN-water, $\phi = 0.60, \alpha = 1$) also shows the buckling and the coexistence of two R$\bar{3}$m phases under shear as discussed below. The shear stress controlled flow curve is obtained by varying the shear stress in the range 700 Pa to 1700 Pa in 30 logarithmic steps with a waiting time of 50 s per data point [Fig. \ref{F8}(a)]. The unaligned diffraction pattern transforms to an aligned pattern at $\sigma$ = 1133 Pa [Fig. \ref{F8}(b)]. Four diffuse arcs appear on the (101) Bragg ring of the R$\bar{3}$m. For $\sigma$ = 1400 Pa (corresponding $\dot{\gamma} \sim$ 1 s$^{-1}$), six arcs are observed on the (101) Bragg ring as well as the azimuthal spread of the arcs become smaller [Fig. \ref{F8}(c)]. For $\sigma$ = 1600 Pa (corresponding $\dot{\gamma} \sim$ 10 s$^{-1}$), two R$\bar{3}$m phases with different lattice parameters appear [Fig. \ref{F8}(d)]. The two sets of lattice parameters ($a = 8.13$ nm, $c = 13.80$ nm) and ($a = 7.69$ nm, $c = 13.80$ nm) [Supplemental Material, Table S7] show lower values compared to the equilibrium lattice parameters ($a = 8.30$ nm, $c = 14.10$ nm). We have followed the temporal evolution of the X-ray diffraction pattern during the shear stress relaxation measurement at 1 s$^{-1}$ [Fig. \ref{F9}(a)]. Again the sharp isotropic Bragg rings in the quiescent state [Fig. \ref{F9}(b)] transform to an aligned diffraction pattern at $t \sim$ 70 s [Fig. \ref{F9}(c)]. The oriented (003) reflection in the vorticity direction indicates the transition to the a-oriented state. At $t = 75$ s, eight diffuse scattering arcs lying on (101), (012) Bragg rings are observed [Fig. \ref{F9}(d)] which is due to buckling of the R$\bar{3}$m. At $t = 80$ s [Fig. \ref{F9}(e)], additional Bragg rings start appearing indicating the onset of another structural transition. The X-ray diffraction pattern obtained at $t = 550$ s shows a stable pattern in the steady-state having a few additional reflections. Eight arcs are seen on different Bragg rings. Considering all the Bragg rings, the diffraction pattern can be indexed to two R$\bar{3}$m phases with different lattice parameters [Supplemental Material, Table S8]. Due to high viscosity, the concentrated R$\bar{3}$m phase was not studied in Couette geometry to avoid breaking of the Couette glass geometry.

\section*{Conclusions}

Using techniques such as Rheo-SAXS, Rheo-SALS and microscopy, the present study has shown two striking consequences of shear flow on the random and ordered mesh phases: (i) A random mesh phase subjected to shear develops in-plane hexagonal ordering of pores and gets locked into a 3D structure to form R$\bar{3}$m phase; (ii) Shearing of R$\bar{3}$m results in the isomorphic twinning and buckling of bilayers out of the shear plane with the subsequent coexistence of two R$\bar{3}$m phases with different lattice parameters. Shear is known to predominantly induce or anneal defects in colloidal crystals though occasionally they can order textural defects that occur in polydomain samples \cite{chatterjee2012formation}. We argue that under shear the NEPT from the random mesh phase to the ordered mesh phase is likely to be a more general feature of the mesh phases and is expected in surfactant-water systems where the random mesh phase and the ordered mesh phase are present adjacent to each other in the equilibrium phase diagram. A crucial point to be noted is that the equilibrium phase transition from a random mesh phase to an ordered mesh phase occurs with decreasing water content where the correlation length of the pores increases with the surfactant's weight fraction ($\phi$). Intriguingly, here in our studies, the lamellar d-spacing decreases under shear only at higher values of $\phi$ where the NEPT appears, providing us with a clue to the origin of this NEPT. We propose that the decrease in the bilayer separation under shear increases the strength of the interaction potential which locks them into a 3D lattice due to the in-plane modulations. This decrease in the lamellar periodicity can arise possibly from the squeezing out of the water from the adjacent bilayers of crystallites, or alternately due to the increase in the average size of the pores. 

Further, rheology of the rhombohedral mesh phase probed in two complimentary experimental geometries (Plate-Plate geometry, and Couette geometry) indicate a shear alignment followed by isomorphic twinning and buckling of the bilayers out of the shear plane. For the ordered mesh phase sheared in a Couette geometry, the buckling and twinning which is observed more towards the stator near the centre of the gap, could be related to the plastic deformation at high shear rates, akin to those observed in soft colloidal crystals \cite{taheri2013shear}. We cannot rule out completely the other possibility that the bilayers of the ordered mesh phase under shear roll into the multi-lamellar cylinders arranged in hexagonal or tetragonal array giving rise to 6 or 8 arcs in the (101), (012) Bragg rings. In that case, large fluctuations in the viscosity even after a long time (Fig. \ref{F7}, \ref{F9}) can be due to the changing orientation of these cylinders under shear. However, the decrease in viscosity does not support the formation of cylinders since rolling up of lamellae into cylinders usually increases the resistance to flow and results in increased viscosity \cite{zipfel2001cylindrical}. A more detailed study would be required to probe the structure and underlying mechanism of the buckling phenomena in ordered mesh phases. 

In order to draw a functional relationship between the timescale of NEPT and the shear rate, one has to check whether the transition is strain-controlled or shear rate-controlled process. If it is a strain-controlled process, one expects inverse relationship between the timescale and the shear rate \cite{fujii2016kinetics}. Otherwise for the shear rate/stress-controlled process, a power-law dependence of the timescale with the applied shear rate/stress is expected. However, a detailed study is required to establish this, a future direction of our study. We hope that our experimental studies will motivate quantitative theoretical understanding of shear-induced transitions in concentrated mixed surfactant systems.

\section*{ASSOCIATED CONTENT}
\textbf{Supporting Information}\\
The Supporting Information is attached after page 25 which includes; estimation of a single defect's volume, shear-study of L$_\alpha^D$ of CPC-SHN-water system, equilibrium SAXS and SALS study of the R$\bar{3}$m, Rheo-SALS during the flow curve of R$\bar{3}$m, supplemental Figures and Tables.

\section*{Conflicts of interest}

There are no conflicts to declare.

\section*{Acknowledgements}

A.K.S. thanks Department of Science and Technology (DST), India for the support under Year of Science Professorship. R. K. thanks DST, India for the Ramanujan Fellowship. V.R. thanks the Council for Scientific and Industrial Research (CSIR), India for the Senior Research Fellowship. P.K.B. thank University Grants Commission (UGC), India for the Senior Research Fellowship. We thank DST, India for financial assistance through DST-DESY Project (I-20140281) to use the Synchrotron Beam-time. We acknowledge DESY (Hamburg, Germany), a member of the Helmholtz Association HGF, for the provision of experimental facilities. Parts of this research were carried out at PETRA III and we thank Dr. Michael Sprung, Dr. Alexey Zozulya, and Eric Stellamanns for assistance in using the P10 beamline.


\newpage

\clearpage
\renewcommand{\thefigure}{S\arabic{figure}}
\renewcommand{\thetable}{S\arabic{table}}
\setcounter{figure}{0}
\section{Supplemental Material}

\subsection{Equilibrium phase diagrams of CTAB-SHN-water and CPC-SHN-water}
\begin{figure*}[htbp]
	\centering
	\includegraphics[width=1\textwidth]{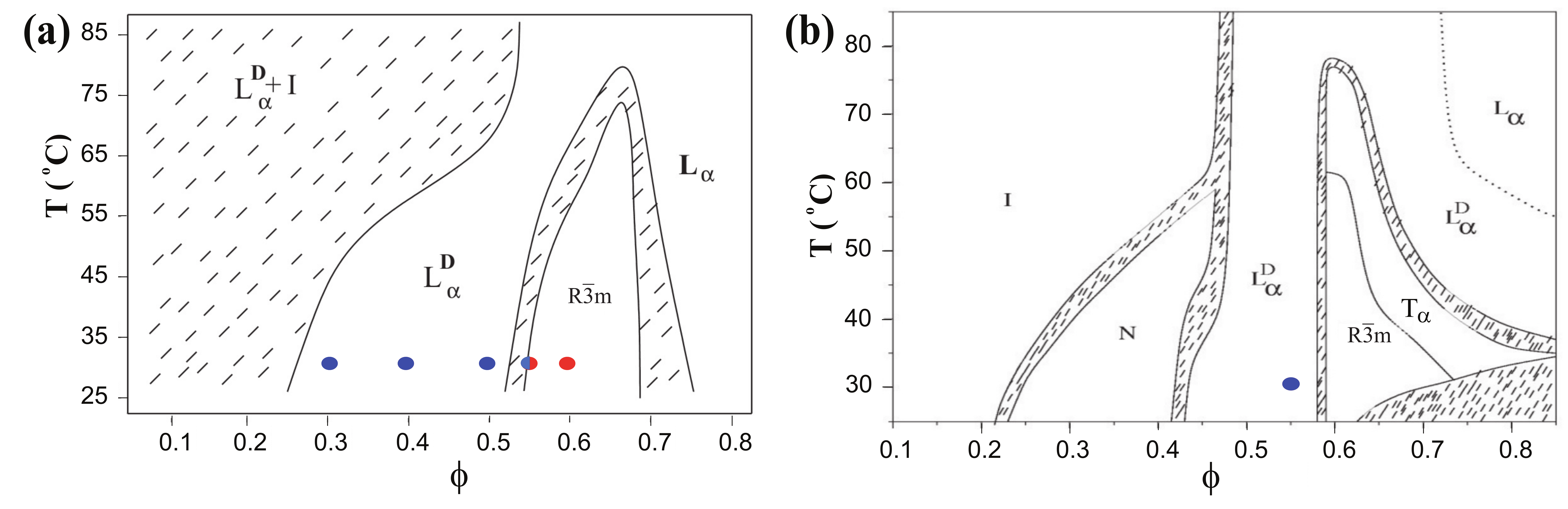}
	\caption{(a) Equilibrium phase diagram showing total weight fraction of surfactant + SHN ($\phi$) vs temperature (T) of CTAB-SHN-water at equimolar ratio ($\alpha =1$). (b) Similar phase diagram of CPC-SHN-water at molar ratio $\alpha =0.5$. L$^{D}_{\alpha}$, I, R$\bar{3}$m, L$_{\alpha}$, N, T$_{\alpha}$ denote the random mesh, isotropic, rhombohedral mesh, lamellar, nematic, tetragonal mesh phases, respectively. The blue and the red dot marks show the compositions studied by us. The phase diagrams are adopted from \cite{krishnaswamy2005phase}, and \cite{gupta2013controlling}, respectively.} 
	\label{CtabCpcPhaseDiagram}
\end{figure*}

Equilibrium phase diagram of the CTAB-SHN-water system ($\alpha =1$) is shown in Fig. \ref{CtabCpcPhaseDiagram}(a). Figure \ref{CtabCpcPhaseDiagram}(b) shows the equilibrium phase diagram of the CPC-SHN-water system ($\alpha =0.5$). At $30^{\circ}$ C, the L$^{D}_{\alpha}$ phase is observed for 0.25 $< \phi <$ 0.5 in CTAB-SHN-water system, and for 0.43 $< \phi <$ 0.57 in CPC-SHN-water system. At $30^{\circ}$ C, the R$\bar{3}$m phase is observed for 0.5 $< \phi <$ 0.7 in CTAB-SHN-water system, and for 0.57 $< \phi <$ 0.75 in CPC-SHN-water system.

\begin{figure*}[htbp]
	\centering
	\includegraphics[width=1\textwidth]{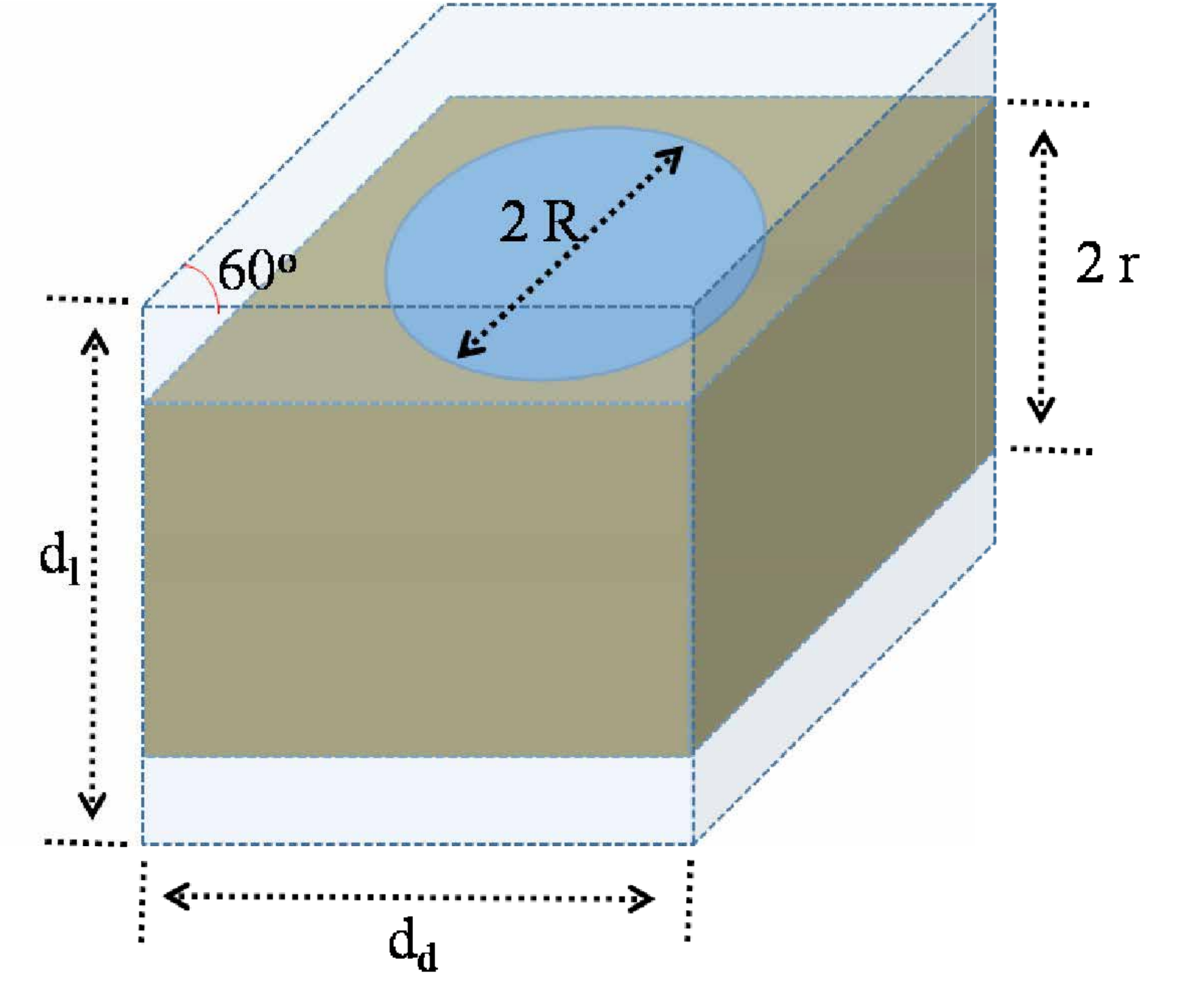}
	\caption{Right rhombic prism of sides d$_d$, d$_d$ with angle 60$^{\circ}$ and height d$_l$. Surfactant layer of thickness 2r has a cylinder of diameter 2R at the centre filled with water. Top and bottom space out side the surfactant is filled with water.} 
	\label{SIFig4}
\end{figure*}

\subsection{Estimation of a single pore's volume of L$^{D}_{\alpha}$ and R$\bar{3}$m using the right rhombic prism model [Fig. \ref{SIFig4}]} 
Using the surfactant weight fraction $\phi$ as defined in the main text, assuming the same density for the surfactant and SHN i.e $\rho_s$ in the bilayers, it can be shown [Fig. \ref{SIFig4}] that if V$_w$ is the total volume occupied by water and V$_s$ is the total volume occupied by surfactant+SHN then we have,
\begin{eqnarray}
	&\Rightarrow& \frac{V_w}{V_s} = \frac{f_\rho}{\psi}; \quad \textrm{where} \quad f_\rho = \frac{\rho_s}{\rho_w}\quad \textrm{and} \quad \psi = \frac{W_{surfactant+SHN}}{W_{water}} = \frac{\phi}{1-\phi}\\
	&\Rightarrow& \frac{(\sqrt{3}/2)d_{d}^{2}(d_{l}-2r)+\pi R^{2} 2r}{((\sqrt{3}/2)d_{d}^{2}-\pi R^{2} )2r} = \frac{f_\rho}{\psi}\\
	&\Rightarrow& \pi R^{2} 2r(f_{\rho}+\psi) = (\sqrt{3}/2)d_{d}^{2}(2rf_{\rho} -(d_{l}-2r)\psi)\\
	&\Rightarrow& \textrm{pore volume} \sim \pi R^{2} 2r = \frac{(\sqrt{3}/2)d_{d}^{2}(2rf_{\rho} -(d_{l}-2r)\psi)}{(f_{\rho}+\psi)}\\
\end{eqnarray}
Table \ref{T1} shows the estimated pore volume of L$^{D}_{\alpha}$ and R$\bar{3}$m for different surfactant concentration, using the measured in-plane pore correlation length ($d_d$) and the bilayer periodicity ($d_l$) of L$^{D}_{\alpha}$ phase or using the measured lattice parameters of R$\bar{3}$m phase at equilibrium.

\begin{table}[h] 
	\caption{\textbf{Estimation of volume of the in-plane water filled pore for different $\phi$. Taken value of r = 2.1 nm (from literature) and f$_\rho$ = 1.03 (measured).}}
	\begin{tabular}{  c  c  c  c  c  c  }
		\hline
		\hline
		system & $\phi$ & the phase & d$_d$ or a (nm) & d$_l$ or c/3 (nm) & pore volume (nm$^3$)\\ \hline
		CTAB-SHN-Water & 0.40 & L$^{D}_{\alpha}$ & 7.68 & 7.05 & 72.60 \\
		& 0.50 & L$^{D}_{\alpha}$ & 7.67 & 5.49 & 76.19 \\
		& 0.53 & R$\bar{3}$m      & 8.95 & 5.44 & 94.06 \\
		& 0.60 & R$\bar{3}$m      & 8.30 & 4.70 & 84.32 \\ \hline
		CPCl-SHN-Water & 0.55 & L$^{D}_{\alpha}$ & 6.50 & 5.13 & 51.90 \\ 
		\hline
		\hline
	\end{tabular}
	\label{T1}
\end{table}  

\clearpage

\subsection{SAXS diffractograms showing L$^{D}_{\alpha}$ to R$\bar{3}$m transition, and the reversibility} 

\begin{figure*}[htbp]
	\centering
	\includegraphics[width=1\textwidth]{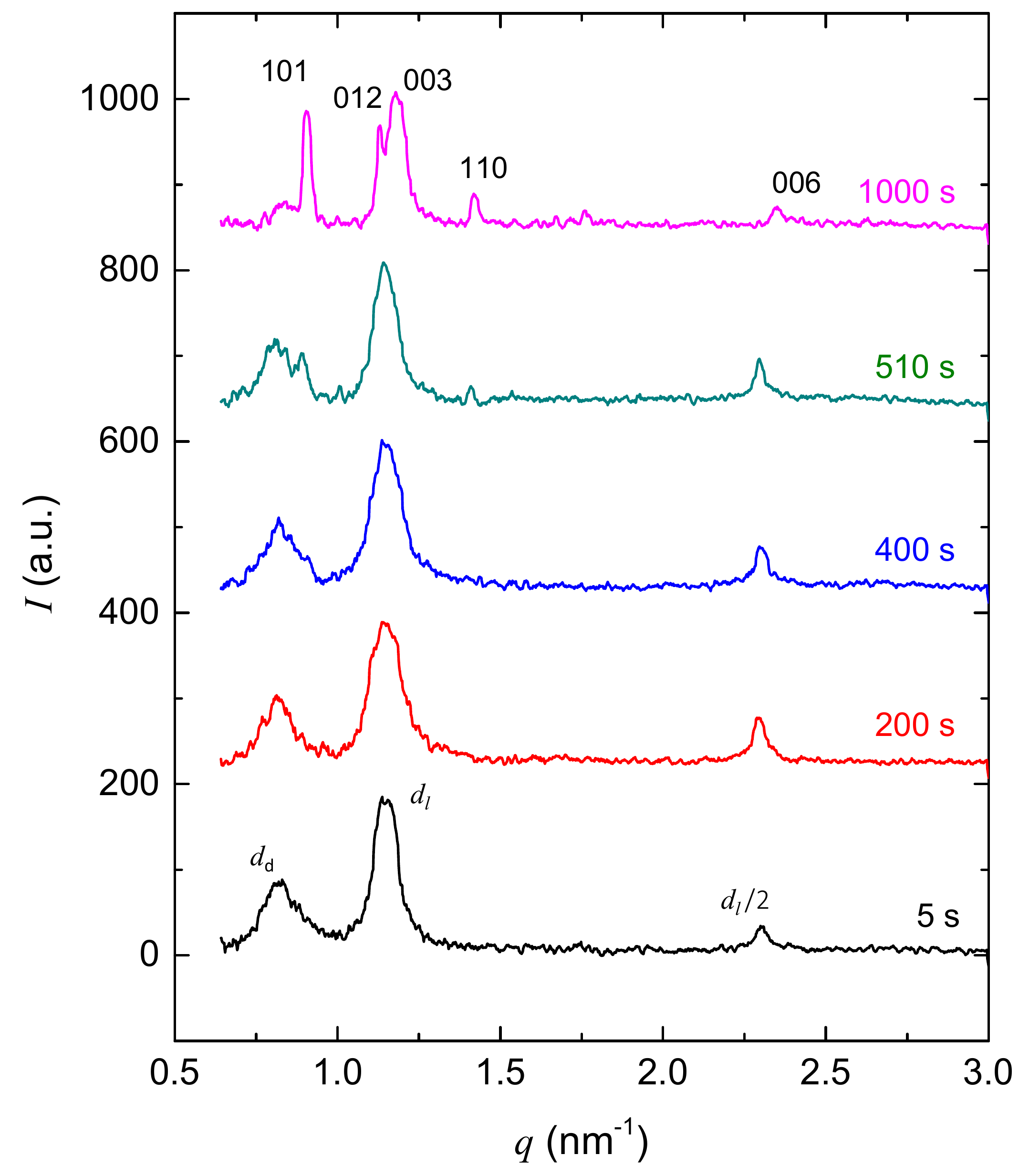}
	\caption{Temporal evolution of the SAXS diffractogram during the random mesh phase (L$^{D}_{\alpha}$) to rhombohedral mesh phase (R$\bar{3}$m) transition at a shear rate = 50 s$^{-1}$ using PP geometry, for CTAB-SHN-water system ($\phi = 0.5$, $\alpha = 1$). Initial Bragg peaks corresponding to the lamellar d-spacing ($d_l$), the diffuse scattering peak due to the in-plane correlation of the nano-pores ($d_d$), and a few Bragg peaks of the shear-induced R$\bar{3}$m are indicated.} 
	\label{IqtCTABLDtoR3m}
\end{figure*}

\begin{figure*}
	\includegraphics[width=1.0\textwidth]{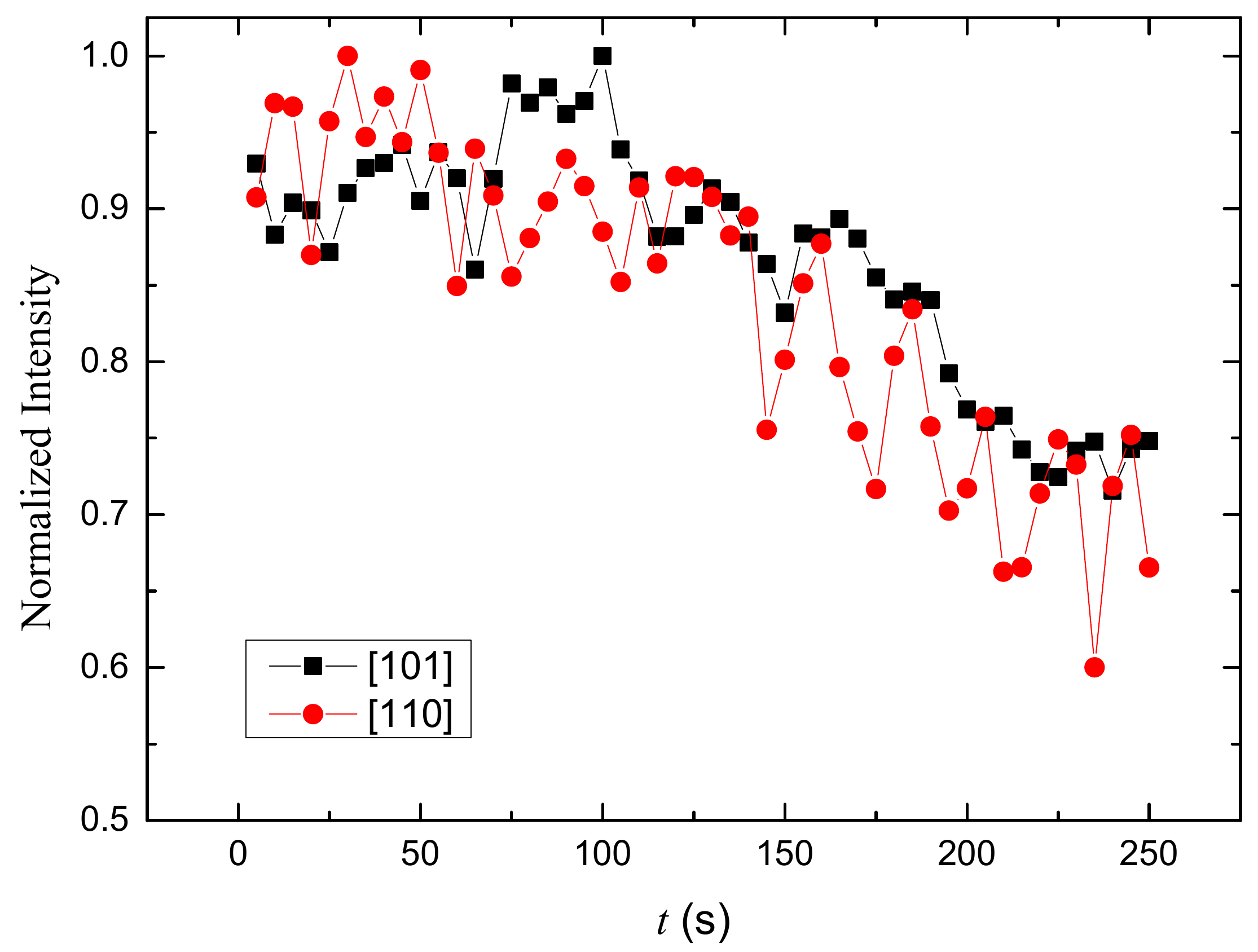}
	\caption{Decrease of the normalized intensities of [101], and [110] Bragg peaks (normalized with respect to intensity at the time of stopping the shear) during the relaxation of the shear induced R$\bar{3}$m phase after stopping the shear using PP geometry, for CTAB-SHN-water ($\phi = 0.5$, $\alpha = 1$).} 
	\label{ItCTABrevLDr3m}
\end{figure*}

For CTAB-SHN-water system ($\phi = 0.5$, $\alpha = 1$) during the random mesh phase (L$^{D}_{\alpha}$) to rhombohedral mesh phase (R$\bar{3}$m) transition, the SAXS diffractograms are calculated by azimuthal integration of intensity with respect to the SAXS pattern centre, and then averagred over no of pixels, considered during azimuthal integration [Fig. \ref{IqtCTABLDtoR3m}]. After stopping the shearing experiment the SAXS pattern was recorded for 250 s in order to check the reversibility of the transition. Figure \ref{ItCTABrevLDr3m} shows the normalized intensities of [101], and [110] Bragg peaks (normalized with respect to intensity at the time of stopping the shear) during the relaxation of the shear induced R$\bar{3}$m phase. 

\clearpage

\subsection{Shear-induced L$^{D}_{\alpha}$ phase to R$\bar{3}$m non-equilibrium phase transition (NEPT) in CPC-SHN-water system, and the reversibility}
\begin{figure*}
	\includegraphics[width=1.0\textwidth]{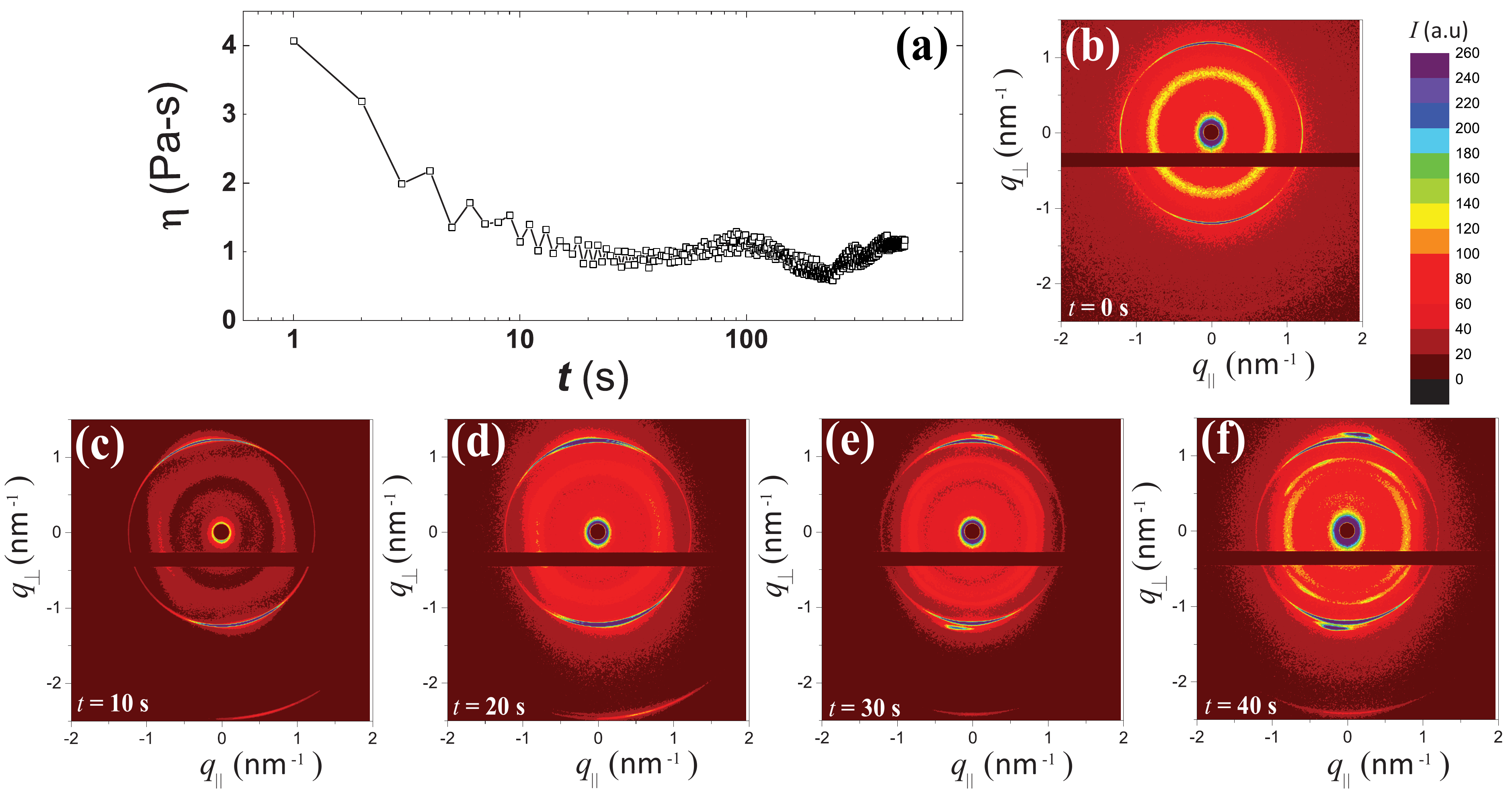}
	\caption{With PP geometry, shear-induced NEPT from L$^{D}_{\alpha}$ phase to R$\bar{3}$m phase is observed during stress relaxation measurement with CPC-SHN-water system ($\alpha = 0.5$, $\phi = 0.55$) at $\dot{\gamma} = 50$ s$^{-1}$. (a) $\eta$ vs $t$ plot and the corresponding X-ray diffraction patterns at (b) $t = 0$ s, (c) $t = 10$ s, (d) $t = 20$ s, (e) $t = 30$ s, (f) $t = 40$ s are shown. The intensity color scale (shown in the right corner) is the same for all the patterns in this figure.} 
	\label{SIF3}
\end{figure*}

\begin{figure*}
	\includegraphics[width=1.0\textwidth]{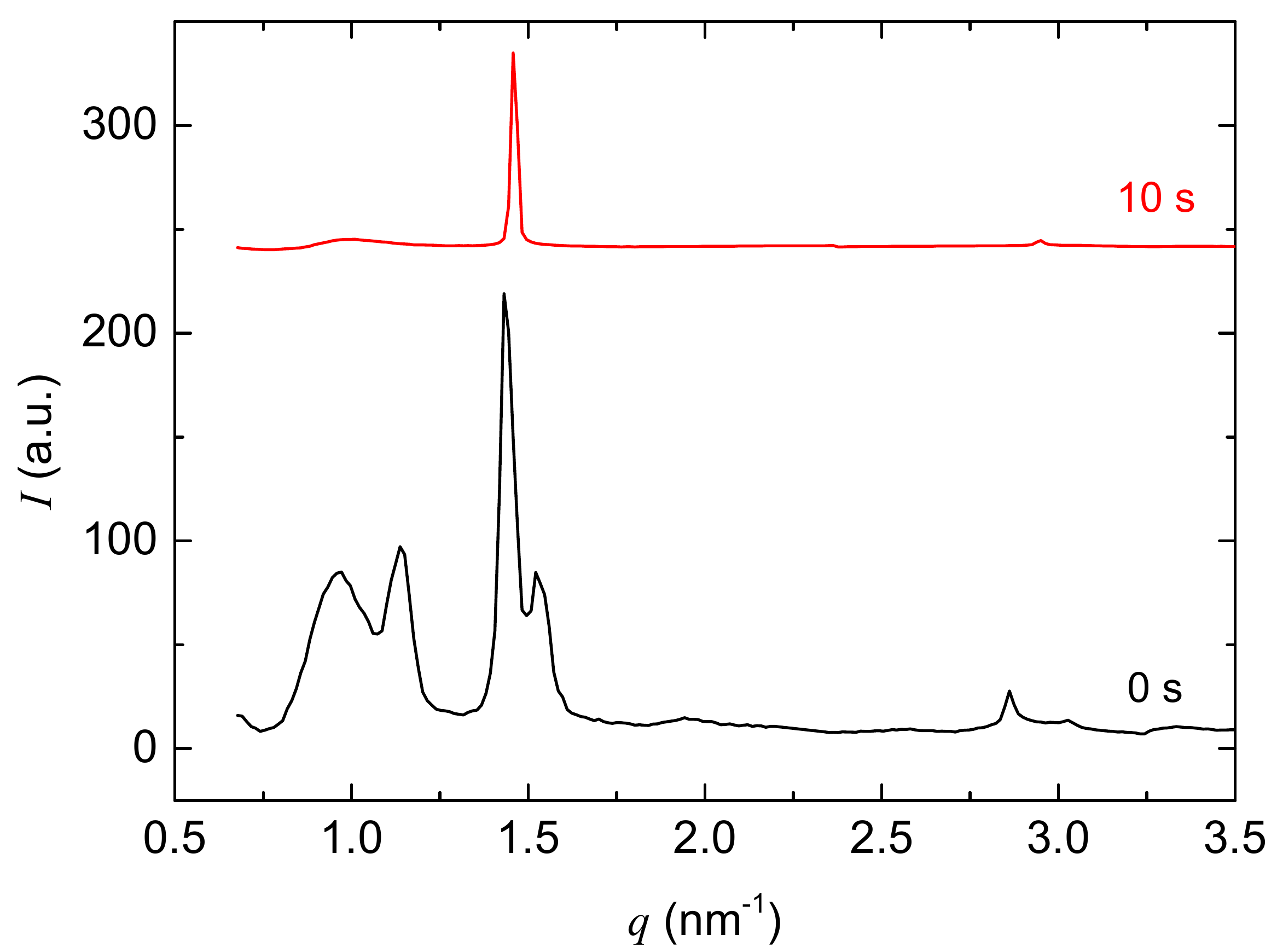}
	\caption{Time relaxation of the shear induced rhombohedral mesh phase (R$\bar{3}$m), obtained by shearing the random mesh phase (L$^{D}_{\alpha}$) at $\dot{\gamma}$ = 50 s$^{-1}$, for CPC-SHN-water ($\alpha = 0.5$, $\phi = 0.55$).} 
	\label{IqtCPCrevLDr3m}
\end{figure*}

With PP geometry, shear-induced L$^{D}_{\alpha}$ phase to R$\bar{3}$m non-equilibrium phase transition (NEPT) in the CPC-SHN-water system ($\alpha = 0.5$, $\phi = 0.55$) is observed during the stress relaxation at $\dot{\gamma} = 50$ s$^{-1}$ [Fig. \ref{SIF3}(a)]. In equilibrium, as shown in Fig. \ref{SIF3}(b), the X-ray diffraction pattern reveals an unoriented lamellar with $d$-spacing of 5.13 nm, coexisting with the diffuse peak from nano-pores with 6.50 nm liquid-like average correlation length. A nearly a-oriented diffraction pattern is obtained at $t \sim 10$ s, with the average orientation of lamellar peaks along $\mathbf{q_{\bot}}$ [Fig. \ref{SIF3}(c)]. The azimuthal spread of the lamellar peak decreases upon further shearing and at $t \sim 20$ s [Fig. \ref{SIF3}(d)], a sharp peak comes up after the diffuse peak indicating 3D ordering of nano-pores getting established under shear similar to the observation in the case of CTAB-SHN-water system (discussed in the main text). After $t = 30$ s, other higher order peaks start appearing [Fig. \ref{SIF3}(e)]. All the sharp reflections obtained at $t \sim 40$ s [Fig. \ref{SIF3}(f)], can be indexed to two R$\bar{3}$m phases with lattice parameters $a1 = 8.39$ nm; $c1 = 14.79$ nm; $a2 = 8.27$ nm; $c2 = 14.28$ nm [Table \ref{T2}]. The SAXS pattern was recorded for 250 s in order to check the reversibility of the transition after stopping the shearing experiment. Figure \ref{IqtCPCrevLDr3m} shows the time relaxation of the shear induced R$\bar{3}$m phase.

\begin{table}[h!]
	\caption{Indexing the X-ray diffraction pattern of the shear-induced R$\bar{3}$m phase [Fig. \ref{SIF3}(f)], obtained by shearing the L$^{D}_{\alpha}$ phase (CPC-SHN-water system, $\alpha$ = 0.5, $\phi$ = 0.55) at $\dot{\gamma}$ = 50 s$^{-1}$. Peaks are fitted to set of two R$\bar{3}$m phases with the calculated unit cell parameters as $a1 = 8.39$ nm, $c1 = 14.79$ nm ($1^{st}$ R$\bar{3}$m) and $a2 = 8.27$ nm, $c2 = 14.28$ nm ($2^{nd}$ R$\bar{3}$m).}
	\begin{tabular}{  c  c  c  c  c  c  c }
		\hline
		\hline
		& \multicolumn{2}{|c||}{1$^{st}$ R$\bar{3}$m} & \multicolumn{2}{c|}{2$^{nd}$ R$\bar{3}$m} & \\ \cline{2-5} 
		$d_{obs}$ & $hkl$ & $d_{cal}$ & $hkl$ & $d_{cal}$ & error  & intensity \\ 
		(nm)   &     &    (nm)   &     &    (nm)   & ($\%$)	&          \\ 
		\hline
		
		6.52 & 101 & 6.52 &     &      & 0   	  & strong \\
		6.40 &     &      & 101 & 6.40 & 0	    & strong\\
		5.06 & 012 & 5.18 & 012 & 5.06 & 2.3, 0 & very strong\\
		4.93 & 003 & 4.93 &     &      & 0      & very strong\\
		4.76 &     &      & 003 & 4.76 & 0      & very strong\\
		3.20 & 113 & 3.19 & 202 & 3.20 & 0.3, 0 & weak \\
		3.10 &     &      & 113 & 3.12 & 0.6    & weak \\
		2.53 & 006 & 2.47 & 122 & 2.53 & 2.4, 0 & weak \\ 
		\hline
		\hline
	\end{tabular}
	\label{T2}
\end{table}

\clearpage
\subsection{Equilibrium study of the rhombohedral mesh phase using Rheo-SALS and Rheo-SAXS setups}
\begin{figure*}[h]
	\centering
	\includegraphics[width=1.0\textwidth]{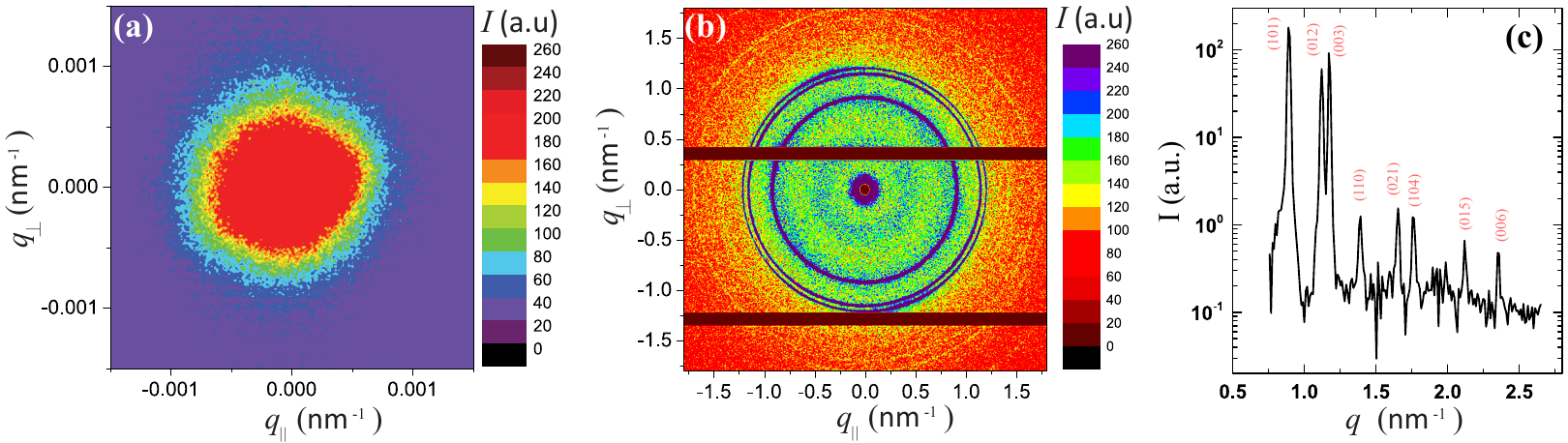}
	\caption{Equilibrium patterns from the rhombohedral phase formed by CTAB-SHN-Water ternary system at $30^\circ$ C: (a) Small angle light scattering (SALS) pattern (weight fraction $\phi=0.53$, and $\alpha=1$) and (b) small angle X-ray scattering (SAXS) pattern (weight fraction $\phi=0.65$, and $\alpha=1$) are shown. These patterns were captured while the samples were resting between the rheometer plates. (c) Intensity vs wave vector ($q$) plot of the SAXS shows the positions of the isotropic rings in $q$-space. Fitted lattice panes (having R$\bar{3}$m symmetry) are indicated. The intensity color scales are shown after the respective patterns.}
	\label{F16}
\end{figure*}

Figure \ref{F16} shows the SALS and the SAXS patterns from the ordered mesh phase at equilibrium just before the rheology measurements. The equilibrium patterns are azimuthally isotropic centering the central beam spot and independent of the geometry in use, a typical signature of randomly oriented crystalline domains. The positions of the SAXS rings in $q$-space can be indexed to a rhombohedral unit cell having the space group symmetry R$\bar{3}$m.

\clearpage
\subsection{Rheo-SALS during the shear stress controlled flow curve measurement with R$\bar{3}$m phase}

\begin{figure*}[h]
	\centering
	\includegraphics[width=1.0\textwidth]{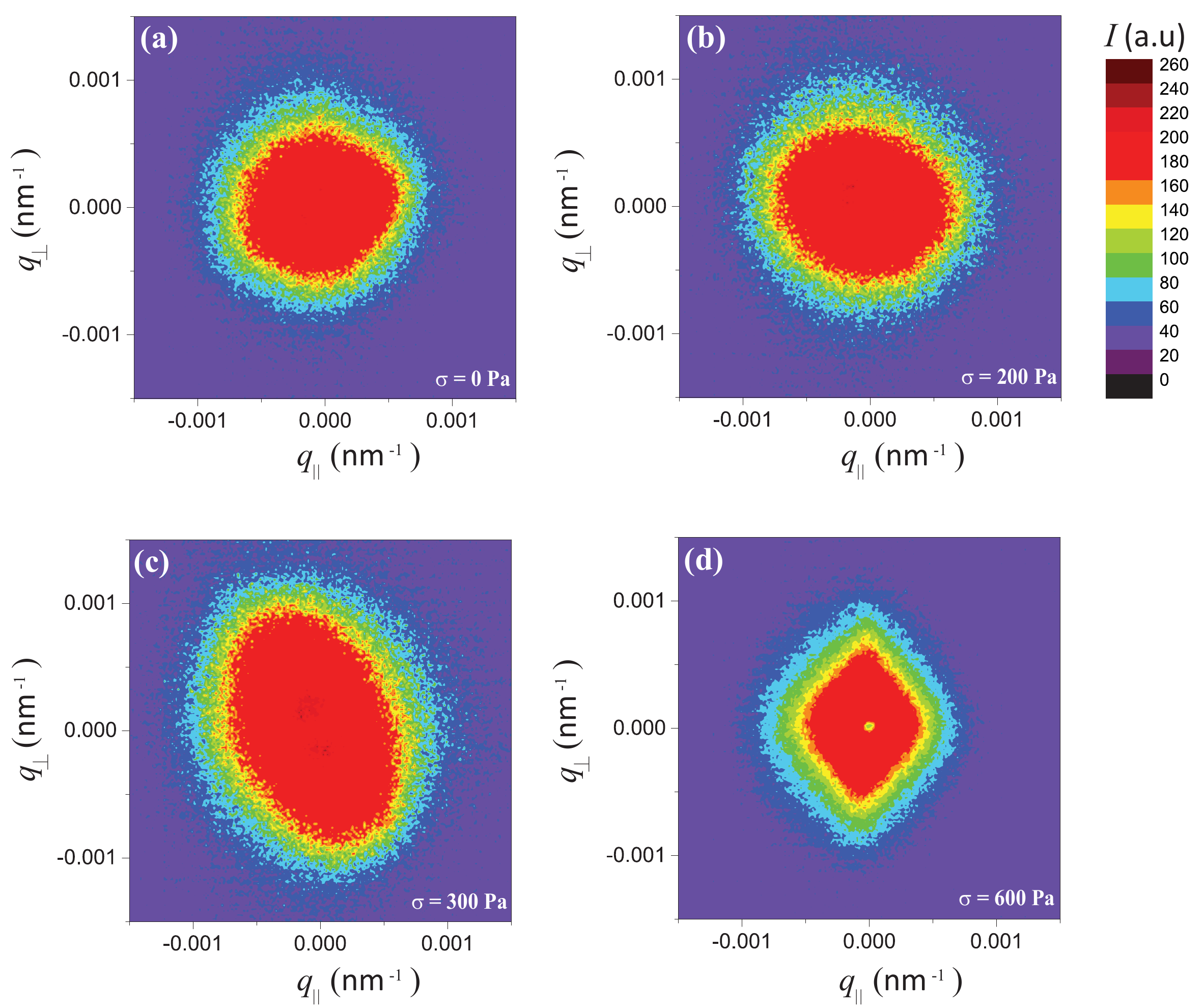}
	\caption{SALS patterns obtained during the shear stress controlled flow curve measurement, are shown for different $\sigma$: (a) 0Pa, (b) 200Pa, (c) 300Pa, (d) 600Pa. The intensity color scale (shown in the right corner) is the same for all the patterns in this figure.}
	\label{R3mF3}
\end{figure*}

The rheo-SALS measurements were performed separately in PP glass geometry with VH configuration. Figure \ref{R3mF3} shows the shear stress-controlled flow curve where the shear stress was varied from 100 Pa to 1000 Pa with a waiting time of $200 \:s$ at each point. The unaligned scattering pattern [Fig. \ref{R3mF3}(a)] transforms to a partially aligned at $\sigma = 200 \:Pa$ [Fig. \ref{R3mF3}(b)] and then shows a-oriented bilayers, bilayer planes parallel to the velocity-velocity gradient plane at $300 \:Pa$ [Fig. \ref{R3mF3}(c)]. Interestingly, at $600 \: Pa$ and above, a star-like pattern appears [Fig. \ref{R3mF3}(d)].

\clearpage
\subsection{The lattice parameters of R$\bar{3}$m at different position in the Couette gap}

\begin{figure}[h]
	\centering
	\includegraphics[width=0.6\textwidth]{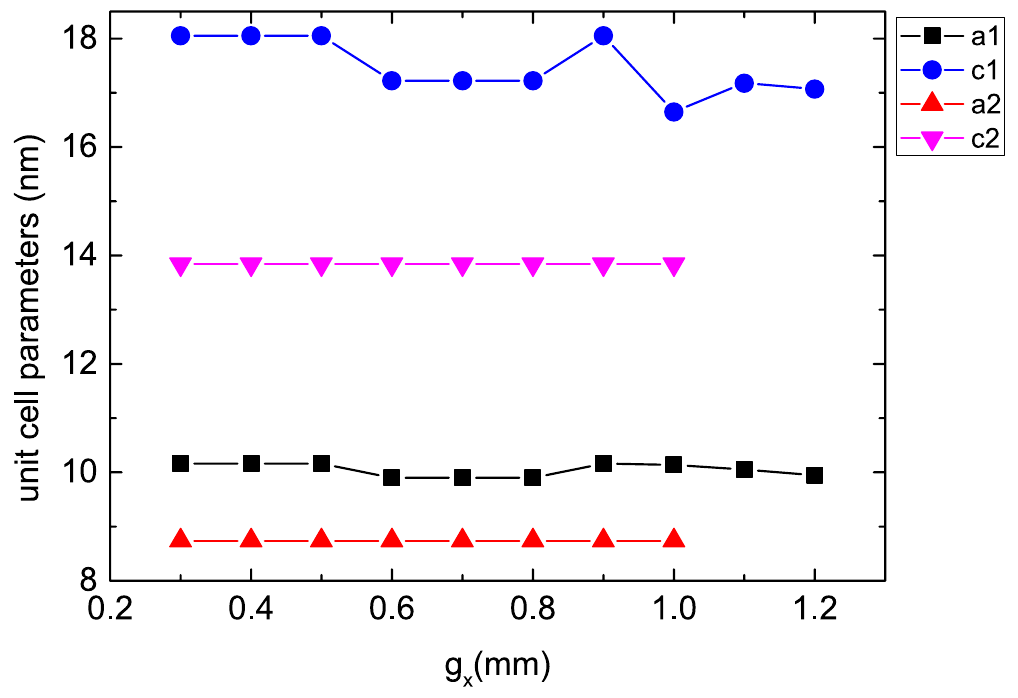}
	\caption{Lattice parameters vs the distance of the x-ray beam from the inner stator cylinder ($g_x$) during the shear rate relaxation of R$\bar{3}$m (CTAB-SHN-water system, $\alpha = 1$, $\phi = 0.53$) for  $\sigma = 100$ Pa, done in Couette geometry. (Patterns are shown in the main text Fig. 7(b)).}
	\label{UnitCellParsOfFig6b}
\end{figure}

The change in the lattice parameters of the R$\bar{3}$m vs the distance of the x-ray beam from the inner stator cylinder of the Couette geometry ($g_x$) is plotted [Fig. \ref{UnitCellParsOfFig6b}], during the shear rate relaxation of R$\bar{3}$m (CTAB-SHN-water system, $\alpha = 1$, $\phi = 0.53$) for  $\sigma = 100$ Pa.

\clearpage

\begin{table}[!ht]
	\caption{Indexing the X-ray diffraction patterns shown in the main text Fig. 4, obtained by shearing the L$^{D}_{\alpha}$ phase (CTAB-SHN-water system, $\alpha$ = 1, $\phi$ = 0.5) at $\dot{\gamma}$ = 50 s$^{-1}$. The average in-plane nano-pore separation ($d_d$) and the lamellar $d$-spacing ($d_l$) are marked. For the patterns at $t = 515$ s and at $t = 850$ s revealing the shear-induced R$\bar{3}$m phase, the calculated $d$-spacings ($d_{cal}$) are obtained using the relation $(1/d)^2 = (4/3)(h^2+hk+k^2)/a^2 + l^2/c^2$ with the condition $-h+k+l = 3n$, where $n$ is an integer. The calculated unit cell parameters $a = 8.68$ nm, $c = 15.93$ nm are same for both.}
	\begin{tabular}{  c  c  c  c  c  c }
		\hline
		\hline
		$t$ & $d_{obs}$       & $hkl$ & $d_{cal}$ & error  & intensity \\
		(s) &   (nm)          &       &    (nm)   & ($\%$) &           \\
		\hline
		0  & 7.67($d_d$) &         &        & 0       & broad \\
		& 5.49($d_l$)        &         &        & 0       & very strong \\
		& 2.76($d_l/2$)      &         &        & 0.5     & strong  \\
		\hline
		
		50  & 7.92($d_d$) &         &        & 0       & broad \\
		& 5.31($d_l$)        &         &        & 0       & very strong \\
		& 2.68($d_l/2$)      &         &        & 0.9     & strong  \\ 
		\hline
		
		515  & 7.79($d_d$)   &         &        &  0   & broad \\
		& 6.80               &  101    & 6.80   &  0   & very strong \\
		& 5.49               &  012    & 5.47   & 0.4  &  strong  \\ 
		& 5.31($d_l$)        &  003    & 5.31   & 0    &  very strong  \\
		& 4.39               &  110    & 4.34   & 1.1  &  weak  \\
		& 2.68($d_l/2$)      &  006    & 2.66   & 0.8  &  strong  \\
		\hline
		
		850  & 7.79($d_d$)   &         &        & 0    & broad \\
		& 6.80               &  101    & 6.80   & 0    & very strong \\
		& 5.49               &  012    & 5.47   & 0.4  &  very strong  \\ 
		& 5.31($d_l$)        &  003    & 5.31   & 0    &  very strong  \\
		& 4.39               &  110    & 4.34   & 1.1  &  strong  \\
		& 3.77               &  021    & 3.66   & 2.9  & weak \\
		& 3.53               &  104    & 3.52   & 0.3  &  strong  \\ 
		& 3.02               &  015    & 2.93   & 3    &  weak  \\
		& 2.66($d_l/2$)      &  006    & 2.66   & 0    &  strong  \\ 
		\hline
		\hline
	\end{tabular}
	\label{TableofF3}
\end{table}

\begin{table}[!ht]
	\caption{Indexing the X-ray diffraction patterns shown in the main text Fig. 5, for different $\sigma$.}
	\begin{tabular}{  c  c  c  c  c  c  c }
		\hline
		\hline
		$\sigma$ & $d_{obs}$ & $hkl$ & $d_{cal}$ & error & intensity & unit cell \\
		(Pa) & (nm) &	 & (nm) & ($\%$) &  & (nm) \\ 
		\hline
		
		0	& 7.00 & 101 & 7.00 & 0 & very strong & $a = 8.95$ \\ 
		& 5.65 & 012 & 5.62 & 0.5 & weak & $c = 16.32$ \\ 
		& 5.44 & 003 & 5.44 & 0 & strong & \\ 
		& 4.47 & 110 & 4.47 & 0 & very strong & \\ 
		& 3.78 & 021 & 3.77 & 0.3 & strong & \\ 
		& 2.81 & 024 & 2.81 & 0 & weak & \\
		& 2.28 & 220 & 2.24 & 1.8 & weak & \\
		\hline
		470	& 7.00 & 101 & 7.00 & 0 & very strong & $a = 8.96$ \\ 
		& 5.61 & 012 & 5.60 & 0.2 & weak & $c = 16.20$ \\ 
		& 5.40 & 003 & 5.40 & 0 & strong & \\ 
		& 4.47 & 110 & 4.48 & 0.2 & very strong & \\ 
		& 3.78 & 021 & 3.77 & 0.3 & strong & \\ 
		& 3.59 & 104 & 3.59 & 0 & strong & \\ 
		& 2.97 & 015 & 2.99 & 0.7 & weak & \\
		& 2.69 & 006 & 2.70 & 0.4 & weak & \\
		\hline
		590	& 7.00 & 101 & 7.00 & 0 & very strong & $a = 9.00$ \\   	
		& 5.61 & 012 & 5.57 & 0.7 & strong & $c = 15.93$ \\ 
		& 5.31 & 003 & 5.31 & 0 & very strong & \\ 
		& 4.47 & 110 & 4.50 & 0.7 & strong & \\ 
		& 2.66 & 006 & 2.66 & 0 & strong & \\ 
		\hline
		660	& 7.00 & 101 & 7.00 & 0 & very strong & $a = 8.98$ \\ 
		& 5.57 & 012 & 5.59 & 0.4 & very strong & $c = 16.05$ \\ 
		& 5.35 & 003 & 5.35 & 0 & very strong & \\ 
		& 4.45 & 110 & 4.49 & 0.9 & very strong & \\ 
		& 3.78 & 021 & 3.78 & 0 & weak & \\ 
		& 3.55 & 104 & 3.57 & 0.6 & strong &  \\
		& 2.96 & 015 & 2.97 & 0.3 & weak & \\
		& 2.67 & 006 & 2.68 & 0.4 & very strong & \\ 
		\hline
		\hline
	\end{tabular}
	\label{TableofF4}
\end{table}

\begin{table}[!ht]
	\caption{Indexing the X-ray diffraction pattern shown in the main text Fig. 6(b) for $g_x = 0.6$ mm. Peaks are fitted to set of two R$\bar{3}$m with calculated unit cell parameters as $a1 = 10.58$ nm, $c1 = 18.60$ nm ($1^{st}$ R$\bar{3}$m) and $a2 = 8.50$ nm, $c2 = 14.25$ nm ($2^{nd}$ R$\bar{3}$m).}
	\begin{tabular}{  c  c  c  c  c  c  c  }
		\hline
		\hline
		& \multicolumn{2}{|c||}{1$^{st}$ R$\bar{3}$m} & \multicolumn{2}{c|}{2$^{nd}$ R$\bar{3}$m} & \\ \cline{2-5} 
		$d_{obs}$ & $hkl$ & $d_{cal}$ & $hkl$ & $d_{cal}$ & error  & intensity  \\ 
		(nm)   &     &    (nm)   &     &    (nm)   & ($\%$)	&           \\ 
		\hline
		
		8.22 & 101 & 8.22 &     &      & 0        & very strong \\
		6.54 & 012 & 6.53 & 101 & 6.54 & 0.2, 0   & very strong \\
		6.20 & 003 & 6.20 &     &      & 0        & very strong \\
		5.22 & 110 & 5.29 &     &      & 1.3      & very strong \\
		5.10 &     &      & 012 & 5.12 & 0.4      & very strong \\
		4.75 &     &      & 003 & 4.75 & 0        & very strong \\
		2.93 & 006 & 3.10 & 113 & 3.17 & 5.5, 7.6 & strong \\
		\hline
		\hline
	\end{tabular}
	\label{TableofF5}
\end{table}

\begin{table}[!ht]
	\caption{Indexing the X-ray diffraction patterns observed at $g_x = 0.8$ mm for different $\sigma$, shown in the main text Fig. 7. For $\sigma =100$ Pa, peaks are fitted to set of two R$\bar{3}$m phases with calculated unit cell parameters as $a1$, $c1$ ($1^{st}$ R$\bar{3}$m) and $a2$, $c2$ ($2^{nd}$ R$\bar{3}$m). Lamellar peaks are marked by $d_l$ and $d_l/2$. Peak intensity vs, s, and w represent very strong, strong, and weak respectively.}
	\begin{tabular}{  c  c  c  c  c  c  c  c  c }
		\hline
		\hline
		&   & \multicolumn{2}{|c||}{1$^{st}$ R$\bar{3}$m} & \multicolumn{2}{c|}{2$^{nd}$ R$\bar{3}$m} & \\ \cline{3-6} 
		$\sigma$ & $d_{obs}$ & $hkl$ & $d_{cal}$ & $hkl$ & $d_{cal}$ & error  & intensity & unit cell \\ 
		(Pa) &  (nm)   &     &    (nm)   &     &    (nm)   & ($\%$)	&          & (nm) \\ 
		\hline
		
		75 & 6.86 & 101 & 6.86 &  &  & 0 & s & $a = 8.84$\\
		& 5.44 & 012 & 5.44 &  &  & 0 & w & $c = 15.48$\\
		& 5.16 & 003 & 5.16 &  &  & 0 & vs & \\
		\hline
		100 & 6.73 & 101 & 6.73 &  &  & 0 & vs & $a1 = 8.65$\\
		& 5.65 &  &  & 101 & 5.65 & 0 & vs & $c1 = 15.36$\\
		& 5.35 & 012 & 5.36 &  &  & 0.19 & vs & $a2 = 7.43$\\
		& 5.12 & 003 & 5.12 &  &  & 0 & vs & $c2 = 11.79$\\
		& 4.42 &  &  & 012 & 4.35 & 1.58 & vs & \\
		& 4.34 & 110 & 4.32 &  &  & 0.46 & w & \\
		& 3.93 &  &  & 003 & 3.93 & 0 & vs & \\
		& 3.76($d_l$) &  &  &  &  &  & vs & \\
		& 3.72 &  &  & 110 & 3.72 & 0 & vs & \\
		& 3.38 & 202 & 3.37 &  &  & 0.30 & vs & \\
		& 2.82 & 015 & 2.84 &  &  & 0.70 & s & \\
		& 2.62 & 006 & 2.56 &  &  & 2.29 & vs & \\
		& 2.54 & 030 & 2.50 &  &  & 1.58 & vs & \\
		& 2.21 &  &  & 015 & 2.21 & 0 & s & \\
		& 1.98 &  &  & 006 & 1.96 & 1.01 & s & \\
		& 1.86($d_l$/2) & 401 & 1.86 &  &  & 0 & w & \\ 
		\hline
		\hline
	\end{tabular}
	\label{TableofF6}
\end{table}

\begin{table}[!ht]
	\caption{Indexing the X-ray diffraction pattern shown in the main text Fig. 8(d). Peaks are fitted to set of two R$\bar{3}$m phases with calculated unit cell parameters as $a1 = 8.13$ nm, $c1 = 13.80$ nm ($1^{st}$ R$\bar{3}$m) and $a2 = 7.69$ nm, $c2 = 13.80$ nm ($2^{nd}$ R$\bar{3}$m).}
	\begin{tabular}{  c  c  c  c  c  c  c }
		\hline
		\hline
		& \multicolumn{2}{|c||}{1$^{st}$ R$\bar{3}$m} & \multicolumn{2}{c|}{2$^{nd}$ R$\bar{3}$m} & \\ \cline{2-5} 
		$d_{obs}$ & $hkl$ & $d_{cal}$ & $hkl$ & $d_{cal}$ & error  & intensity \\ 
		(nm)   &     &    (nm)   &     &    (nm)   & ($\%$) &      \\ 
		\hline
		
		6.27 & 101 & 6.27 &     &      & 0  & very strong \\
		6.00 &     &      & 101 & 6.00 & 0  & very strong \\
		4.82 &     &      & 112 & 4.79 & 0.6& very strong \\
		4.60 & 003 & 4.60 & 003 & 4.60 &0, 0& very strong \\
		4.09 & 110 & 4.06 &     &      & 0.7& weak \\
		3.89 &     &      & 110 & 3.85 & 1.0& weak \\
		3.48 & 021 & 3.41 &     &      & 2.0& strong \\
		3.26 &     &      & 021 & 3.24 & 0.6& strong \\
		3.08 &     &      & 104 & 3.06 & 0.6& strong \\
		2.30 & 006 & 2.30 &     &      & 0  & strong \\ 
		\hline
		\hline	  
		\label{TableofF7}
	\end{tabular}
	\label{T8}
\end{table}

\begin{table}[!ht]
	\caption{Indexing the X-ray diffraction pattern shown in the main text Fig. 9(f). Peaks are fitted to set of two R$\bar{3}$m phases with calculated unit cell parameters as $a1 = 7.76$ nm, $c1 = 13.80$ nm ($1^{st}$ R$\bar{3}$m) and $a2 = 7.38$ nm, $c2 = 13.17$ nm ($2^{nd}$ R$\bar{3}$m).}
	\begin{tabular}{  c  c  c  c  c  c  c }
		\hline
		\hline
		& \multicolumn{2}{|c||}{1$^{st}$ R$\bar{3}$m} & \multicolumn{2}{c|}{2$^{nd}$ R$\bar{3}$m} & \\ \cline{2-5} 
		$d_{obs}$ & $hkl$ & $d_{cal}$ & $hkl$ & $d_{cal}$ & error  & intensity \\ 
		(nm)   &     &    (nm)   &     &    (nm)   & ($\%$) &      \\ 
		\hline
		
		6.04 & 101 & 6.04 &     &      & 0         & very strong \\
		5.75 &     &      & 101 & 5.75 &      0    & very strong \\
		4.83 & 012 & 4.81 &     &      & 0.4,0     & very strong \\
		4.60 & 003 & 4.60 & 012 & 4.59 & 0,0.22    & very strong \\
		4.39 &     &      & 003 & 4.39 & 0         & very strong \\
		3.08 & 104 & 3.07 & 021 & 3.11 & 0.32,0.97 & weak \\ 
		2.29 & 006 & 2.30 & 024 & 2.29 & 0.43, 0   & strong \\ 
		\hline
		\hline
		\label{TableofF8}
	\end{tabular}
\end{table}

\clearpage

\bibliographystyle{ieeetr}
\bibliography{RefMeshPhase}

\begin{thebibliography}{10}

\bibitem{kekicheff1989structure}
P.~Kekicheff and G.~J.~T. Tiddy, ``Structure of the intermediate phase and its
  transformation to lamellar phase in the lithium perfluorooctanoate/water
  system,'' {\em The Journal of Physical Chemistry}, vol.~93, no.~6,
  pp.~2520--2526, 1989.

\bibitem{krishnaswamy2005phase}
R.~Krishnaswamy, S.~K. Ghosh, S.~Lakshmanan, V.~A. Raghunathan, and A.~K. Sood,
  ``{Phase behavior of concentrated aqueous solutions of cetyltrimethylammonium
  bromide (CTAB) and sodium hydroxy naphthoate (SHN)},'' {\em Langmuir},
  vol.~21, no.~23, pp.~10439--10443, 2005.

\bibitem{gupta2013controlling}
S.~P. Gupta and V.~A. Raghunathan, ``Controlling the thermodynamic stability of
  intermediate phases in a cationic-amphiphile--water system with strongly
  binding counterions,'' {\em Physical Review E}, vol.~88, no.~1, p.~012503,
  2013.

\bibitem{cates1990statics}
M.~E. Cates and S.~J. Candau, ``Statics and dynamics of worm-like surfactant
  micelles,'' {\em Journal of Physics: Condensed Matter}, vol.~2, no.~33,
  p.~6869, 1990.

\bibitem{holmes2005bicontinuous}
M.~C. Holmes and M.~S. Leaver, ``Intermediate phases. {In Bicontinuous Liquid
  Crystals (Surfactant Science Series)},'' {\em Lynch, M. L., Spicer, P. T.,
  Eds.; Taylor \& Francis: Boca Raton}, vol.~127, pp.~15--39, 2005.

\bibitem{funari1992microscopy}
S.~S. Funari, M.~C. Holmes, and G.~J.~T. Tiddy, ``{Microscopy, X-ray
  diffraction, and NMR studies of lyotropic liquid crystal phases in the
  C22EO6/water system: a new intermediate phase},'' {\em The Journal of
  Physical Chemistry}, vol.~96, no.~26, pp.~11029--11038, 1992.

\bibitem{leaver2001structural}
M.~Leaver, A.~Fogden, M.~Holmes, and C.~Fairhurst, ``{Structural Models of the
  R$\bar{3}$m Intermediate Mesh Phase in Nonionic Surfactant Water Mixtures},''
  {\em Langmuir}, vol.~17, no.~1, pp.~35--46, 2001.

\bibitem{ghosh2007structure}
S.~K. Ghosh, R.~Ganapathy, R.~Krishnaswamy, J.~Bellare, V.~A. Raghunathan, and
  A.~K. Sood, ``Structure of mesh phases in a cationic surfactant system with
  strongly bound counterions,'' {\em Langmuir}, vol.~23, no.~7, pp.~3606--3614,
  2007.

\bibitem{lucassen1981surface}
E.~H. Lucassen-Reynders, J.~Lucassen, and D.~Giles, ``Surface and bulk
  properties of mixed anionic/cationic surfactant systems in equilibrium
  surface tensions,'' {\em Journal of Colloid and Interface Science}, vol.~81,
  no.~1, pp.~150--157, 1981.

\bibitem{manohar1986origin}
C.~Manohar, U.~R.~K. Rao, B.~S. Valaulikar, and R.~M. Lyer, ``On the origin of
  viscoelasticity in micellar solutions of cetyltrimethylammonium bromide and
  sodium salicylate,'' {\em Journal of the Chemical Society, Chemical
  Communications}, no.~5, pp.~379--381, 1986.

\bibitem{kaler1992phase}
E.~W. Kaler, K.~L. Herrington, A.~K. Murthy, and J.~A.~N. Zasadzinski, ``Phase
  behavior and structures of mixtures of anionic and cationic surfactants,''
  {\em The Journal of Physical Chemistry}, vol.~96, no.~16, pp.~6698--6707,
  1992.

\bibitem{yatcilla1996phase}
M.~T. Yatcilla, K.~L. Herrington, L.~L. Brasher, E.~W. Kaler, S.~Chiruvolu, and
  J.~A. Zasadzinski, ``{Phase behavior of aqueous mixtures of
  cetyltrimethylammonium bromide (CTAB) and sodium octyl sulfate (SOS)},'' {\em
  The Journal of Physical Chemistry}, vol.~100, no.~14, pp.~5874--5879, 1996.

\bibitem{blandamer2000titration}
M.~J. Blandamer, B.~Briggs, P.~M. Cullis, and J.~B. F.~N. Engberts, ``Titration
  microcalorimetry of mixed alkyltrimethylammonium bromide surfactant aqueous
  solutions,'' {\em Physical Chemistry Chemical Physics}, vol.~2, no.~22,
  pp.~5146--5153, 2000.

\bibitem{rand1968x}
R.~P. Rand and V.~Luzzati, ``X-ray diffraction study in water of lipids
  extracted from human erythrocytes: the position of cholesterol in the lipid
  lamellae,'' {\em Biophysical Journal}, vol.~8, no.~1, pp.~125--137, 1968.

\bibitem{hyde2003novel}
S.~Hyde and G.~Schr{\"o}der, ``Novel surfactant mesostructural topologies:
  between lamellae and columnar (hexagonal) forms,'' {\em Current opinion in
  colloid \& interface science}, vol.~8, no.~1, pp.~5--14, 2003.

\bibitem{yang2003rhombohedral}
L.~Yang and H.~W. Huang, ``A rhombohedral phase of lipid containing a membrane
  fusion intermediate structure,'' {\em Biophysical Journal}, vol.~84, no.~3,
  pp.~1808--1817, 2003.

\bibitem{sakamoto2021direct}
K.~Sakamoto, T.~Morishita, K.~Aburai, D.~Ito, T.~Imura, K.~Sakai, M.~Abe,
  I.~Nakase, S.~Futaki, and H.~Sakai, ``Direct entry of cell-penetrating
  peptide can be controlled by maneuvering the membrane curvature,'' {\em
  Scientific reports}, vol.~11, no.~1, pp.~1--9, 2021.

\bibitem{diat1993effect}
O.~Diat, D.~Roux, and F.~Nallet, ``Effect of shear on a lyotropic lamellar
  phase,'' {\em Journal de Physique II}, vol.~3, no.~9, pp.~1427--1452, 1993.

\bibitem{bergenholtz1996formation}
J.~Bergenholtz and N.~J. Wagner, ``{Formation of AOT/brine multilamellar
  vesicles},'' {\em Langmuir}, vol.~12, no.~13, pp.~3122--3126, 1996.

\bibitem{zipfel1999influence}
J.~Zipfel, J.~Berghausen, P.~Lindner, and W.~Richtering, ``Influence of shear
  on lyotropic lamellar phases with different membrane defects,'' {\em The
  Journal of Physical Chemistry B}, vol.~103, no.~15, pp.~2841--2849, 1999.

\bibitem{koschoreck2009multilamellar}
S.~Koschoreck, S.~Fujii, P.~Lindner, and W.~Richtering, ``Multilamellar
  vesicles (“onions”) under shear quench: pathway of discontinuous size
  growth,'' {\em Rheology Acta}, vol.~48, p.~231, 2009.

\bibitem{grobkopf2019shear}
S.~Grobkopf, B.~Tiersch, J.~Koetz, A.~Mix, and T.~Hellweg, ``Shear-induced
  transformation of polymer-rich lamellar phases to micron-sized vesicles,''
  {\em Langmuir}, vol.~35, p.~3048, 2019.

\bibitem{zilman1999undulation}
A.~G. Zilman and R.~Garnek, ``Undulation instability of lamellar phases under
  shear: A mechanism for onion formation?,'' {\em Eur. Phys. J. B}, vol.~11,
  p.~593, 1999.

\bibitem{ramaswamy1992shear}
S.~Ramaswamy, ``Shear-induced collapse of the dilute lamellar phase,'' {\em
  Physical Review Letters}, vol.~69, no.~1, p.~112, 1992.

\bibitem{cates1989role}
M.~E. Cates and S.~T. Milner, ``Role of shear in the isotropic-to-lamellar
  transition,'' {\em Physical Review Letters}, vol.~62, no.~16, p.~1856, 1989.

\bibitem{raghunathan2012mesh}
V.~A. Raghunathan, ``Mesh phases of surfactant-water systems,'' {\em Journal of
  the Indian Institute of Science}, vol.~88, no.~2, pp.~197--210, 2012.

\bibitem{luzzati1960structure}
V.~Luzzati, H.~Mustacchi, A.~Skoulios, and F.~Husson, ``{La structure des
  collo{\"\i}des d'association. I. Les phases liquide--cristallines des
  syst{\`e}mes amphiphile--eau},'' {\em Acta Crystallographica}, vol.~13,
  no.~8, pp.~660--667, 1960.

\bibitem{kekicheff1991cylinders}
P.~K{\'e}kicheff, ``{From cylinders to bilayers: A structural study of phase
  transformations in a lyotropic liquid crystal},'' {\em Molecular Crystals and
  Liquid Crystals}, vol.~198, no.~1, pp.~131--144, 1991.

\bibitem{fairhurst1997structure}
C.~E. Fairhurst, M.~C. Holmes, and M.~S. Leaver, ``{Structure and morphology of
  the intermediate phase region in the nonionic surfactant C16EO6/water
  system},'' {\em Langmuir}, vol.~13, no.~19, pp.~4964--4975, 1997.

\bibitem{holmes1998intermediate}
M.~C. Holmes, ``Intermediate phases of surfactant-water mixtures,'' {\em
  Current Opinion in Colloid and Interface Science}, vol.~3, no.~5,
  pp.~485--492, 1998.

\bibitem{rathee2013reversible}
V.~Rathee, R.~Krishnaswamy, A.~Pal, V.~A. Raghunathan, M.~Imp{\'e}ror-Clerc,
  B.~Pansu, and A.~K. Sood, ``Reversible shear-induced crystallization above
  equilibrium freezing temperature in a lyotropic surfactant system,'' {\em
  Proceedings of the National Academy of Sciences, USA}, vol.~110, no.~37,
  pp.~14849--14854, 2013.

\bibitem{mendes1997vesicle}
E.~Mendes and S.~V.~G. Menon, ``Vesicle to micelle transitions in surfactant
  mixtures induced by shear,'' {\em Chemical Physics Letters}, vol.~275,
  no.~5-6, pp.~477--484, 1997.

\bibitem{fairhurst1996shear}
C.~E. Fairhurst, M.~C. Holmes, and M.~S. Leaver, ``{Shear Alignment of a
  Rhombohedral Mesh Phase in Aqueous Mixtures of a Long Chain Nonionic
  Surfactant},'' {\em Langmuir}, vol.~12, no.~26, pp.~6336--6340, 1996.

\bibitem{struth2011observation}
B.~Struth, K.~Hyun, E.~Kats, T.~Meins, M.~Walther, M.~Wilhelm, and
  G.~Gr{\"u}bel, ``{Observation of new states of liquid crystal 8CB under
  nonlinear shear conditions as observed via a novel and unique
  rheology/small-angle x-ray scattering combination},'' {\em Langmuir},
  vol.~27, no.~6, pp.~2880--2887, 2011.

\bibitem{goulian1995shear}
M.~Goulian and S.~T. Milner, ``Shear alignment and instability of smectic
  phases,'' {\em Physical Review Letters}, vol.~74, no.~10, p.~1775, 1995.

\bibitem{escalante2000shear}
J.~Escalante, M.~Gradzielski, H.~Hoffmann, and K.~Mortensen, ``Shear-induced
  transition of originally undisturbed lamellar phase to vesicle phase,'' {\em
  Langmuir}, vol.~16, no.~23, pp.~8653--8663, 2000.

\bibitem{samuels1971small}
R.~J. Samuels, ``Small-angle light scattering from optically anisotropic
  spheres and disks. theory and experimental verification,'' {\em Journal of
  Polymer Science Part A-2: Polymer Physics}, vol.~9, no.~12, pp.~2165--2246,
  1971.

\bibitem{wunenburger2001oscillating}
A.~Wunenburger, A.~Colin, J.~Leng, A.~Arn{\'e}odo, and D.~Roux, ``Oscillating
  viscosity in a lyotropic lamellar phase under shear flow,'' {\em Physical
  Review Letters}, vol.~86, no.~7, p.~1374, 2001.

\bibitem{bruinsma1992shear}
R.~Bruinsma and Y.~Rabin, ``Shear-flow enhancement and suppression of
  fluctuations in smectic liquid crystals,'' {\em Physical Review A}, vol.~45,
  no.~2, p.~994, 1992.

\bibitem{silmore2020buckling}
K.~S. Silmore, M.~Strano, and J.~W. Swan, ``Buckling{,} crumpling{,} and
  tumbling of semiflexible sheets in simple shear flow,'' {\em Soft Matter},
  p.~doi:10.1039/D0SM02184A, 2021 Accepted Article.

\bibitem{hamley2001structure}
I.~W. Hamley, ``Structure and flow behaviour of block copolymers,'' {\em
  Journal of Physics: Condensed Matter}, vol.~13, no.~33, p.~R643, 2001.

\bibitem{chatterjee2012formation}
S.~Chatterjee and S.~L. Anna, ``Formation and ordering of topological defect
  arrays produced by dilatational strain and shear flow in smectic-a liquid
  crystals,'' {\em Physical Review E}, vol.~85, no.~1, p.~011701, 2012.

\bibitem{taheri2013shear}
S.~M. Taheri, S.~Rosenfeldt, S.~Fischer, P.~B{\"o}secke, T.~Narayanan,
  P.~Lindner, and S.~F{\"o}rster, ``Shear-induced macroscopic “siamese”
  twins in soft colloidal crystals,'' {\em Soft Matter}, vol.~9, no.~35,
  pp.~8464--8475, 2013.

\bibitem{zipfel2001cylindrical}
J.~Zipfel, F.~Nettesheim, P.~Lindner, T.~D. Le, U.~Olsson, and W.~Richtering,
  ``Cylindrical intermediates in a shear-induced lamellar-to-vesicle
  transition,'' {\em EPL (Europhysics Letters)}, vol.~53, no.~3, p.~335, 2001.

\bibitem{fujii2016kinetics}
S.~Fujii and Y.~Yamamoto, ``Kinetics of the orientation transition in the
  lyotropic lamellar phase,'' {\em Journal of Biorheology}, vol.~30, no.~1,
  pp.~27--33, 2016.

\end{thebibliography}

\end{document}